\documentclass[prd,aps,twocolumn,superscriptaddress,showpacs,tightenlines,floatfix,nofootinbib,amssymb]{revtex4-1}
\usepackage{graphicx}
\usepackage{epstopdf}

\newcommand{\beqn}{\begin{eqnarray}}
\newcommand{\eeqn}{\end{eqnarray}}
\newcommand{\eq}[1]{(\ref{#1})}
\newcommand{\M}{PLSM${}_q$}

\newcommand{\cZ}{{\cal Z}}
\newcommand{\dd}{{\mathrm d}}
\newcommand{\dD}{{\cal D}}
\newcommand{\Tr}{{\mathrm{Tr}\,}}
\newcommand{\Z}{{\mathbb{Z}}}
\newcommand{\Dirac}{\rlap {\hspace{-0.5mm} \slash} D}
\newcommand{\dirac}{\rlap {\hspace{-1.2mm} \slash} \partial}
\def\bbbone{{\mathchoice {\rm 1\mskip-4mu l} {\rm 1\mskip-4mu l}
{\rm 1\mskip-4.5mu l} {\rm 1\mskip-5mu l}}}

\begin{document}

\title{Phase diagram of hot QCD in an external magnetic field: \\
possible splitting of deconfinement and chiral transitions}

\author{Ana J\'ulia Mizher}\email{anajulia@if.ufrj.br}
\affiliation{Instituto de F\'isica, Universidade Federal do Rio de Janeiro,
Caixa Postal 68528, Rio de Janeiro, RJ 21941-972, Brazil}
\affiliation{Department of Physics and Astronomy, Vrije Universiteit Amsterdam,
De Boelelaan1081, NL-1081HV Amsterdam, the Netherlands}
\author{M.~N.~Chernodub}\email{maxim.chernodub@lmpt.univ-tours.fr}\thanks{\\ on leave from ITEP, Moscow, Russia.}
\affiliation{Laboratoire de Math\'ematiques et Physique Th\'eorique
Universit\'e Fran\c{c}ois-Rabelais, F\'ed\'eration Denis Poisson - CNRS,
Parc de Grandmont, Universit\'e de Tours, 37200 France}
\affiliation{Department of Physics and Astronomy, University of Gent, Krijgslaan 281, S9, B-9000 Gent, Belgium}
\author{Eduardo S. Fraga}\email{fraga@if.ufrj.br}
\affiliation{Instituto de F\'isica, Universidade Federal do Rio de Janeiro,
Caixa Postal 68528, Rio de Janeiro, RJ 21941-972, Brazil}

\begin{abstract}
The structure of the phase diagram for strong interactions becomes richer in the
presence of a magnetic background, which enters as a new control parameter
for the thermodynamics.
Motivated by the relevance of this physical setting for current and future high-energy
heavy ion collision experiments and for the cosmological QCD transitions, we use
the linear sigma model coupled to quarks and to Polyakov loops as an effective theory
to investigate how the chiral and the deconfining transitions are affected, and present
a general picture for the temperature--magnetic field phase diagram. We compute and
discuss each contribution to the effective potential for the approximate order parameters,
and uncover new phenomena such as the paramagnetically-induced breaking of
global $\mathbb{Z}_3$ symmetry, and possible splitting of deconfinement and chiral transitions
in a strong magnetic field.
\end{abstract}


\maketitle

\section{Introduction}

The phase diagram of QCD has a very rich structure, and keeps unveiling new, sometimes
unexpected phases of strong interactions as one plays with thermodynamic control parameters
such as the temperature, chemical potentials and quark masses (for recent results, see e.g.
Ref. \cite{CPOD2009}). A particular external control parameter that is often overlooked, however,
is the magnetic field. Since quarks and charged hadrons couple to the magnetic field and
neutral hadrons do not, this control parameter can affect the phase structure of QCD in a
nontrivial fashion. Moreover, it is present in most physical systems exhibiting deconfinement
and chiral symmetry restoration. The example that was considered to be the most spectacular
is from astrophysics, provided by magnetars \cite{magnetars}. Besides, magnetic fields are
expected to be strong and play an important role in structure formation in the early universe,
during the epochs of the electroweak and the QCD primordial phase transitions \cite{Schwarz:2003du}.
However, it was recently found that non-central high-energy heavy ion collisions might
generate the most intense observed magnetic fields,
much stronger than in magnetars,
reaching values of $B\sim 10^{19}$ Gauss,
which corresponds to $eB\sim 6 m_{\pi}^{2}$, where $e$ is the fundamental charge and $m_{\pi}$
is the pion mass.
These magnetic fields are
short-lived for very high energies
but play an important role in possible experimental signatures of strong CP violation and the phenomenon
of the chiral magnetic
effect~\cite{Kharzeev:1998kz,Kharzeev:2007jp,Skokov:2009qp,Fukushima:2010fe}.

In this paper we present a general picture for the temperature--magnetic field phase diagram
of QCD, restricting our analysis to a vanishing baryonic chemical potential. In particular,
we investigate how the chiral and the deconfining transitions are affected,
and uncover new phenomena such as the {\it paramagnetically-induced breaking} of $\Z_3$.
We also demonstrate the possible splitting of the deconfinement and chiral transition in an
external magnetic field.

For this purpose, we adopt the linear sigma model coupled to quarks with two flavors, $N_{f}=2$,
{\it and} to Polyakov loops
(\M)
under the effect of a magnetic background field as an effective theory for the thermodynamics of
QCD.
The linear sigma model coupled to quarks \cite{GellMann:1960np}
has been widely used to describe different aspects of the chiral transition, such as thermodynamic
properties and the nonequilibrium phase conversion process
\cite{quarks-chiral,ove,Scavenius:1999zc,Caldas:2000ic,Scavenius:2000qd,
Scavenius:2001bb,paech,Mocsy:2004ab,Aguiar:2003pp,Schaefer:2006ds,
Taketani:2006zg,Fraga:2004hp}. It has also been combined to Polyakov loops to include
confinement \cite{Dumitru:2000in,Mocsy:2003tr,Megias:2004hj,Herbst:2010rf}. However, as will be clear
in the next section, we perform this generalization in a different fashion, which is also more
natural in the presence of an external magnetic field.

\begin{figure}[!thb]
\vskip 3mm
\begin{center}
\includegraphics[width=85mm,clip=true]{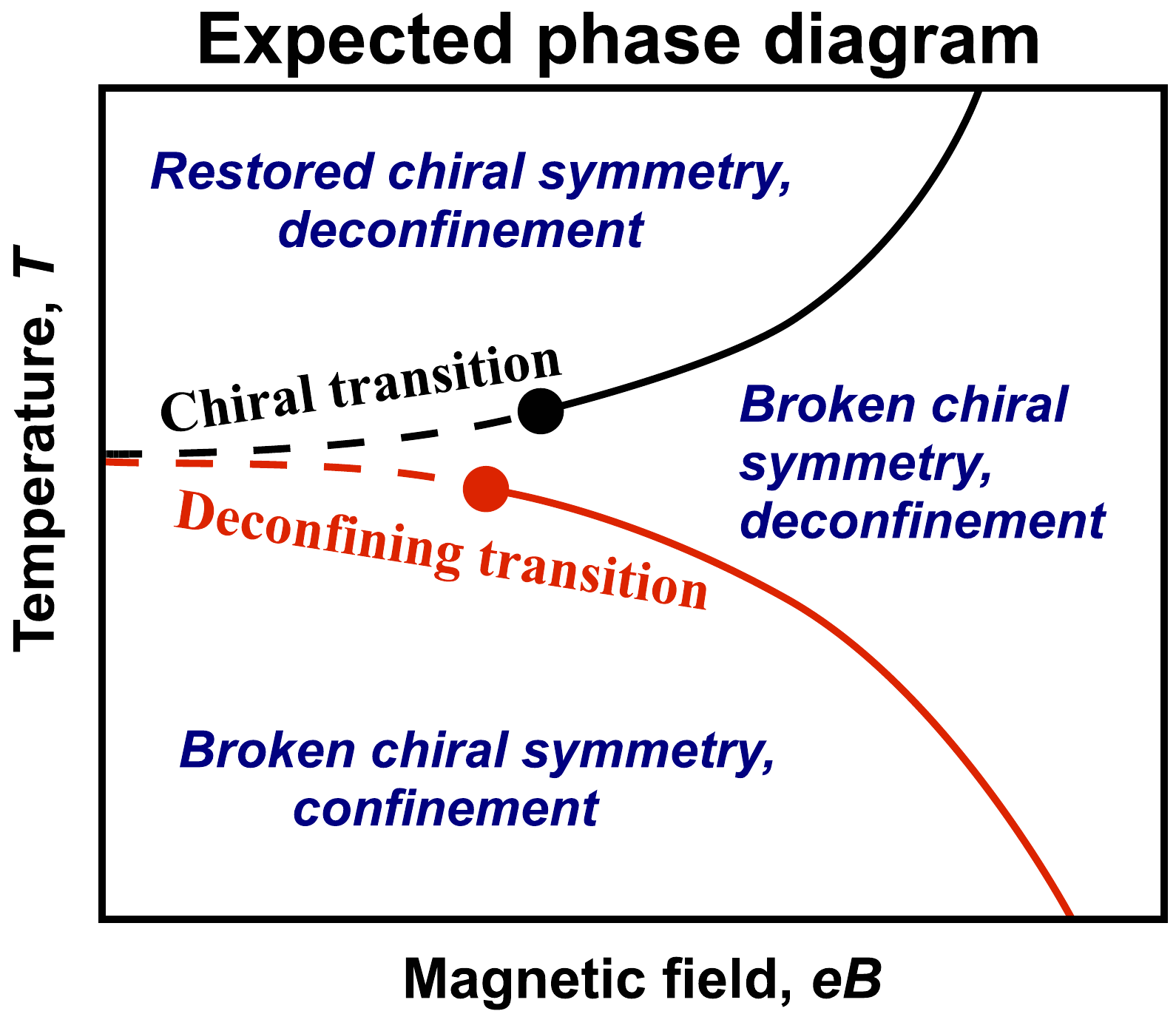}
\end{center}
\vskip -3mm
\caption{Expected magnetic field--temperature phase diagram of QCD.
The thick lines indicate first-order transitions, the filled circles are the (second-order) endpoints
of these lines, and the thin dashed lines stand for the corresponding crossovers. A new phase with
broken chiral symmetry and deconfinement appears at high magnetic fields.}
\label{fig:expected}
\end{figure}

In our effective model, \M,
we compute and discuss each one-loop contribution to the effective potential for the approximate order parameters of the chiral and
deconfining transitions, given by the chiral condensate and the expectation value of the Polyakov
loop, respectively. We treat the external magnetic field, that plays the role of a thermodynamic
control parameter, as a constant and uniform field, and make no assumption on its intensity.
In Fig.~\ref{fig:expected} we show a cartoon of the temperature--magnetic field phase diagram
one would expect from previous results for the limits of strong and weak fields, for chiral and
deconfining transitions treated separately, as discussed below.

Modifications in the vacuum of strong interactions by the presence of an external magnetic
field have been investigated previously within different frameworks, mainly using effective
models \cite{Klevansky:1989vi,Gusynin:1994xp,Babansky:1997zh,Klimenko:1998su,
Semenoff:1999xv,Goyal:1999ye,Hiller:2008eh,Ayala:2006sv,Boomsma:2009yk},
especially the NJL model \cite{Klevansky:1992qe}, and chiral perturbation
theory \cite{Shushpanov:1997sf,Agasian:1999sx,Cohen:2007bt}, but also resorting to the
quark model \cite{Kabat:2002er} and certain limits of QCD \cite{Miransky:2002rp}.
Most treatments have been concerned with vacuum modifications by the magnetic field,
though medium effects were considered in a few cases, as e.g. in the study of the stability
of quark droplets under the influence of a magnetic field at finite density and zero
temperature, with nontrivial effects on the order of the chiral transition \cite{Ebert:2003yk}.
Magnetic effects on the dynamical quark mass \cite{Klimenko:2008mg}, as well as  magnetized
chiral condensates in a holographic description of chiral symmetry breaking \cite{holographic},
were also considered.
Lattice simulations were also used to demonstrate an enhancement of the chiral condensate
and the appearance of chiral magnetization of the chirally broken vacuum
under the influence of a strong magnetic field~\cite{lattice}.
Ring-diagram corrections to the effective potential in a weak magnetic
field were considered in Ref. \cite{ring}. Recent applications of quark matter under strong magnetic
fields to the physics of magnetars using the NJL model can be found in Ref. \cite{Menezes:2008qt,Menezes:2009uc}.

The case of the thermal quark-hadron transition in a magnetic field was studied recently
in Ref.~\cite{Agasian:2008tb}. It was found that the critical temperature and the latent heat
are both diminished in the presence of the magnetic field,
and that the first-order deconfining turns into a crossover at sufficiently strong magnetic fields.
On the other hand, it has been shown in Ref. \cite{Fraga:2008qn} that the chiral transition can be dramatically modified
in the presence of a {\it strong} magnetic field. In particular, the smooth crossover found for
vanishing baryonic densities can be turned into a first-order phase transition if the
magnetic field is sufficiently high, the threshold being close to fields that can be achieved
in current and future heavy ion collision experiments. For the chiral transition --
on the contrary
to the case of the deconfining transition of Ref. \cite{Agasian:2008tb} -- the critical temperature
seems to increase with the magnetic field, thus favoring a first-order behavior, which goes in
line with the theoretical expectation on the role played by magnetic fields in the
nature of phase transitions \cite{landau-book}.
On the other hand, at very weak magnetic field both confinement and the chiral transitions should again
be crossovers~\cite{Aoki:2006we}.
This justifies our preliminary approximate cartoon of Fig.~\ref{fig:expected}.

More recently, the chiral magnetic effect was also studied in the context of the PNJL
model \cite{Fukushima:2010fe}. Here we propose a different effective theory for
the study of the phase structure of QCD in the presence of an external magnetic field,
the linear sigma model coupled to quarks and to the Polyakov loop,
and compute
the $T-eB$  phase diagram in the presence and,
separately,
in the absence of vacuum corrections.
The contribution from these corrections is very large, though the issue of their relevance
or necessity is non-trivial in this context as also in different ones, such as the Higgs potential
in the electroweak theory and the Walecka model in high-density nuclear
physics \cite{FTFT-books}. We choose to present results in both scenarios and address
the issue in the end. Nevertheless, we believe that lattice results will be crucial in setting
the most appropriate effective theory approach. The \M\  can, of course, also be used to
study chiral symmetry breaking and deconfinement in the absence of a magnetic field.
We present a few results in this case but, keeping our focus on the phase structure in the
presence of a magnetic background, leave a thorough study for a future publication~\cite{next}.

The paper is organized as follows. In Section~\ref{sec:model} we provide a description of the
effective theory that allows us to study the influence of an external magnetic field on confining
and chiral properties of QCD, the linear sigma model coupled to quarks and to the Polyakov
loop (\M). We discuss the quark, chiral and confining sectors, the total free energy and the
connection to various observables. In Section~\ref{sec:free-energy} we compute all one-loop
contributions to the effective potential, discussing vacuum and thermal corrections in detail.
Section~\ref{sec:results} contains our results for the effective potential and
a discussion of the temperature--magnetic field phase diagram. Section \ref{sec:conclusion}
contains our conclusions and outlook.

\section{Effective theory}
\label{sec:model}

\subsection{Quark, chiral and confining variables}

The effective field theory that we denote by \M\ has three types of variables:
the quark field $\psi(x)$, the chiral field $\phi(x)$ and the Polyakov loop
variable $L(x)$.

The confining properties of QCD are described by the
complex-valued Polyakov loop variable
$L$, which plays the role of an approximate order parameter for the confining
phase transition in QCD. The Polyakov loop is a singlet part of the
$3 \times 3$ matrix $\Phi$ (``untraced Polyakov loop'') which belongs to
the adjoint representation of the $SU(3)$ color group, i.e.:
\beqn
L(x) = \frac{1}{3} \Tr \Phi(x)\,,
\quad
\Phi = {\cal P} \exp \Bigl[i \int\limits_0^{1/T}
\dd \tau A_4(\vec x, \tau) \Bigr]\,,
\label{eq:L}
\eeqn
where $A_4 = i A_0$ is the matrix-valued temporal component of the Euclidean
gauge field $A_\mu$ and the symbol ${\cal P}$ denotes path ordering. The
integration takes place over compactified imaginary time $\tau$.

The expectation value of the Polyakov loop $L$ is an {\it exact} order parameter
of the color confinement in the limit  of infinitely massive quarks:
\beqn
\mbox{Confinement}:\quad
\left\{
\begin{array}{llll}
\langle L \rangle  & = & 0 \quad , \quad & \mbox{low $T$}   \\
\langle L \rangle  & \neq & 0 \quad , \quad & \mbox{high $T$}
\end{array}
\right.
\label{eq:L:phases}
\eeqn

The chiral features of the model are encoded in the dynamics of the $O(4)$
chiral field
\beqn
\phi =(\sigma,\vec{\pi})\,,
\qquad
\vec{\pi} = (\pi^{+},\pi^{0},\pi^{-})\,,
\label{eq:phi}
\eeqn
where $\vec{\pi}$ is the isotriplet of the pseudoscalar pion fields and
$\sigma$ is the chiral scalar field which plays the role of an approximate
order parameter of the chiral transition. The expectation value of the
field $\sigma$ is an exact order parameter in the chiral limit, in which
quarks and pions are massless degrees of freedom:
\beqn
\mbox{Chiral symmetry}:\quad
\left\{
\begin{array}{llll}
\langle \sigma \rangle  & \neq & 0 \quad , \quad & \mbox{low $T$}   \\
\langle \sigma \rangle  & = & 0 \quad ,\quad & \mbox{high $T$}
\end{array}
\right.
\label{eq:sigma:phases}
\eeqn

The up and down fields of the constituent quarks are grouped together
into the doublet fermion field
\beqn
\psi=
\left(
\begin{array}{c}
u\\
d
\end{array}
\right)\,,
\eeqn
which plays a central role in our discussion. The fermion field
couples two other variables, the Polyakov loop $L$ and the chiral field
$\phi$ to each other, thus linking confining and chiral properties
together. The quarks $\psi$ are also coupled to the external magnetic field
since the $u$ and $d$ quarks are electrically charged. Since the quarks are
also coupled to the Polyakov loop and the chiral field $\phi$, it is clear
that the external magnetic field will affect the chiral dynamics {\it and}
the confining properties of the model. Thus, this model allows us to investigate
the influence of the external magnetic field on color confinement and chiral
symmetry breaking simultaneously.

We can write the Lagrangian of \M\ as follows:
\beqn
{\cal L} = {\cal L}_q(\overline{\psi},\psi,\phi,L) +
           {\cal L}_\phi(\phi) +
           {\cal L}_L(L)\,.
\label{eq:L:full}
\eeqn
%

\subsection{Quark sector}

The first part of the full Lagrangian (\ref{eq:L:full}),
\beqn
{\cal L}_q =
 \overline{\psi} \left[i \Dirac - g(\sigma +i\gamma _{5}
 \vec{\tau} \cdot \vec{\pi} )\right]\psi\,,
\label{eq:L:psi}
\eeqn
describes the constituent quarks,
which interact with the meson fields $\phi$, the Abelian gauge field $a_\mu$
and the $SU(3)$ gauge field $A_\mu$ via the covariant derivative:
\beqn
\Dirac = \gamma^{\mu} D^{(q)}_\mu\,,
\qquad
D^{(q)}_\mu = \partial _{\mu} - i Q\, a_\mu - i A_\mu\,.
\label{eq:D}
\eeqn

The Abelian gauge field describes the influence of the external magnetic field
$\vec B$ aligned along the third direction, $B_i = B \delta_{i3}$, for
convenience:
\beqn
a_\mu = (a^0,\vec a) = (0, - B y,0,0)\,,
\eeqn
and the electric charges of the quarks are defined by the following matrix:
\beqn
Q \equiv
\left(
\begin{array}{cc}
q_u & 0 \\
0 & q_d \\
\end{array}
\right)
=
\left(
\begin{array}{cc}
+ \frac{2}{3}e & 0 \\
0 & - \frac{e}{3} \\
\end{array}
\right)\,,
\eeqn
where $e$ is the elementary electron charge.

The $SU(3)$ gauge field $A_\mu$ represents a nontrivial background due to the
Polyakov loop~\eq{eq:L}. It is convenient to diagonalize the untraced Polyakov
loop $\Phi$. In this gauge the field $A_4$, which enters the quark Lagrangian
\eq{eq:L:psi}, is diagonal:
\beqn
A_4 = t_3\,  A_4^{(3)} + t_8\, A_4^{(8)}\,,
\eeqn
where $t_{3,8} = \lambda_{3,8}/2$ are the Cartan generators of the $SU(3)$ gauge
group. The $SU(3)$ gauge field is prescribed to take constant values, being linked
to the Polyakov loop as follows:
\beqn
\Phi & = & \exp\left[ i \left(\, t_3\,  \frac{A_4^{(3)}}{T}
+ \, t_8\, \frac{A_4^{(8)}}{T} \right)\right]
\nonumber\\
& = &
{\rm diag}~ (e^{i \varphi_1}, ~e^{i \varphi_2}, ~e^{i \varphi_3}) \, ,
\label{eq:Phi}
\eeqn
with
\beqn
\begin{array}{lll}
\varphi_1 \equiv \frac{A^{11}_4}{T} & = & \frac{1}{T} \bigl(\frac{1}{2} A_4^{(3)} +
\frac{1}{2 \sqrt{3}} A_4^{(8)} \bigr)\,,
\\
\varphi_2 \equiv \frac{A^{22}_4}{T} & = & \frac{1}{T} \bigl(- \frac{1}{2} A_4^{(3)} +
\frac{1}{2 \sqrt{3}} A_4^{(8)} \bigr) \,, \\
\varphi_3 \equiv \frac{A^{33}_4}{T} & = & - \frac{1}{T} \frac{1}{\sqrt{3}} A_4^{(8)}\,.
\end{array}
\label{eq:phis}
\eeqn
%

\subsection{Chiral Lagrangian}
\label{chiral_lagrangian}

The chiral part of the \M\ Lagrangian is given by the second term of
Eq.~\eq{eq:L:full}, that can be written in the Minkowski space-time
as follows:
\beqn
{\cal L}_\phi(\sigma,\vec\pi)  =
\frac{1}{2}(\partial _{\mu}\sigma \partial ^{\mu}\sigma + \partial _{\mu}
\pi^0 \partial ^{\mu}\pi^0)
\nonumber\\
+ D^{(\pi)}_\mu \pi^+ D^{(\pi)\mu} \pi^-
- V_\phi(\sigma ,\vec{\pi})\,.
\label{eq:L:phi}
\eeqn
Here we have introduced the charged $\pi^\pm$ and neutral $\pi^0$ mesons,
respectively,
\beqn
\pi^\pm = \frac{1}{\sqrt{2}} (\pi^1 \pm i \pi^2)\,, \qquad \pi^0 = \pi^3\,.
\eeqn
The electric charges of the $\pi^\pm$-mesons are $\pm e$, so that the
covariant derivative in Eq.~\eq{eq:L:phi} reads
\beqn
D_\mu^{(\pi)} = \partial_\mu + i e a_\mu\,.
\label{eq:D:pi}
\eeqn
This derivative does not involve the Polyakov loop, contrary to the covariant
derivative acting on quarks \eq{eq:D}, since the pions are colorless states.

In Eq.~\eq{eq:L:phi}, $V_\phi$  stands for the potential of the chiral fields.
This potential exhibits both spontaneous and explicit breaking of chiral symmetry:
\beqn
V_\phi(\sigma ,\vec{\pi}) & = & \frac{\lambda}{4}(\sigma^{2}+\vec{\pi}^{2} -
{\it v}^2)^2-h\sigma
\label{eq:V:sigma}\\
& = & \frac{1}{2} m_\sigma^2 \sigma^2 + \frac{1}{2} m_\pi^2 (\pi^0)^2 +
m_\pi^2 \pi^+ \pi^- + \dots \,
\nonumber
\eeqn
In the second line of this equation we explicitly show the quadratic form of the potential
which provides the masses to the meson fields. Here, the masses $m_\sigma$ and $m_\pi$
correspond to the vacuum masses of the sigma and the pion mesons, which are used to fix the
parameters $\lambda$ and $v$ in the classical potential, since $f_\pi$ is also known, as explained
below.

As customary in the linear sigma model, we follow a mean-field analysis in which the
mesonic sector is treated classically whereas quarks represent fast degrees of freedom.
The parameters of the Lagrangian \eq{eq:L:phi} are chosen to match the low-energy
phenomenology of mesons in the absence of magnetic fields and at zero temperature, i.e.
in the vacuum \cite{sigma-model}. Using the notation of Ref. \cite{Scavenius:2000qd},
this implies the following expectation values for the condensates:
$\langle\sigma\rangle = f_{\pi}$ and $\langle\vec{\pi}\rangle = 0$,
with $f_\pi \approx 93\,\mbox{MeV}$ and $m_\pi \approx 138\,\mbox{MeV}$.
The masses in Eq.~\eq{eq:V:sigma} should be considered as parameters of the 
model, which coincide with corresponding meson masses at $T=0$ and $B=0$.

The explicit symmetry  breaking term, as usual, is determined by the PCAC
relation $h = f_{\pi} m_{\pi}^2$, so that $v^2 = f^2_\pi-{m^{2}_{\pi}}/{\lambda}$
and $m^2_\sigma = 2 \lambda f^2_\pi + m^2_\pi$ in the vacuum.
Choosing the value of the quartic interaction $\lambda = 20$, one gets a
reasonable $\sigma$-mass $m_\sigma \approx 600\,\mbox{MeV}$.
The constituent quark mass is given by
$m_q \equiv m_q(\langle\sigma\rangle)= g \langle\sigma\rangle$,
and, choosing $g=3.3$ at $T=0$, one obtains for the constituent quarks in the
vacuum $m_q \approx 310\,\mbox{MeV}$. At low temperatures the quarks are not
excited, and the model~\eq{eq:L:phi} reproduces results from the usual linear
$\sigma$-model without quarks~\cite{FTFT-books}.

If the explicit symmetry breaking term is absent, $h_q = 0$, then the
model~\eq{eq:L:phi} experiences a second-order phase transition~\cite{Pisarski:1983ms}
from the broken symmetry phase to the restored phase at a critical temperature
$T_c=\sqrt{2} v$. The presence of the explicit symmetry breaking term changes the
phase transition into a smooth crossover.

The pion directions play no major role in the process of phase conversion we have in mind,
as was argued in Ref. \cite{Scavenius:2001bb} for the case of the chiral transition. Therefore
in what follows we focus on the sigma direction of the chiral sector. However, the coupling of charged
pions to the magnetic fields might be quantitatively important in a possible development of
charged chiral condensates. This issue, though, makes the
computation technically much more involved and will be addressed in a future publication
\cite{next}.

\subsection{Confining potential}

We choose to couple the Polyakov loop to the linear sigma model in the same
fashion as was successfully implemented in the case of the Nambu--Jona-Lasinio
model in Refs.~\cite{ref:fukushima,ref:ratti07,ref:ratti08}. The confining properties
of the system in our model are accounted for by the third term in Eq.~\eq{eq:L:full},
which describes the Polyakov loop variable~\eq{eq:L}:
\beqn
{\cal L}_L = - V_L(L,T)\,.
\label{eq:L:L}
\eeqn
We treat the Polyakov loop in a mean-field approach, neglecting a kinetic term for
this variable. Although the Polyakov loop is, strictly speaking, a static quantity,
a kinetic term is often added to describe the dynamics in an effective model for
deconfinement \cite{Meisinger:2002kg,Fraga:2006cr}.

Notice that, in our approach, the Polyakov loop variable $L$ is not coupled to the chiral
variables $\phi$ in the bare Lagrangian~\eq{eq:L:L}. The reason is that the chiral variables
are colorless degrees of freedom which should not affect the Polyakov loop at least
in a first approximation.  Of course, the mesons $\phi$ are made of colored quarks
which do feel colored degrees of freedom like $L$. Thus, the interaction between
the chiral and confining variables will inevitably appear later, after the quarks
$\psi$ are integrated out. This approach is not unique. In fact, it is common to couple
the Polyakov loop directly to chiral fields phenomenologically as a way to link
the chiral and deconfining transitions in a generalized Landau-Ginzburg theory,
in which the behavior of an order parameter induces a change in the behavior of non-order
parameters at the transition via the presence of a possible coupling between the
fields \cite{Dumitru:2000in,Mocsy:2003tr,Megias:2004hj}.

The term~\eq{eq:L:L} is essentially the potential term of the Polyakov action in the
pure Yang-Mills theory not coupled to quarks. Quark effects manifest via coupling with
the quark matter field in the first part of our Lagrangian, Eq.~\eq{eq:L:psi}.

The effective potential~\eq{eq:L:L} has the following properties: it should satisfy the
center $\Z_3$ symmetry,
\beqn
\Z_3: \qquad L \to e^{2 \pi n i/3} L\,, \quad n=0,1,2\,,
\label{eq:Z3}
\eeqn
which is realized in the limit of the pure gauge theory (the coupling to dynamical
quarks break this symmetry). According to Eq.~\eq{eq:L:phases}, the
potential~\eq{eq:L:L} should have an absolute minimum at $L = 0$ in the confined phase.
Results from lattice simulations~\cite{ref:Lattice:Tc} show that the deconfining transition
in $SU(3)$ gauge theory takes place at a critical temperature
$T_{SU(3)}\simeq 270\, {\mathrm{MeV}}$. As the system undergoes the phase transition
the single minimum at $L = 0$ of the potential splits into three degenerate minima
labeled by the $\Z_3$ variable. As a consequence, in the deconfined phase
the $\Z_3$ symmetry is spontaneously broken, and $\langle L \rangle \neq 0$.

Following Ref.~\cite{ref:ratti08} we use a specific form for the phenomenological potential
of the Polyakov loop:
\beqn
\label{eq:V}
& & \frac{V_L(L,T)}{T^4} =-\frac{1}{2}a(T)\,L^*L \\
& & + b(T)\,\ln\left[1-6\,L^*L+4\left({L^*}^3+L^3\right) - 3\left(L^*L\right)^2\right]\,,
\nonumber
\eeqn
with
\beqn
a(T) & = & a_0 + a_1\left(\frac{T_0}{T}\right) +a_2\left(\frac{T_0}{T}\right)^2\,, \\
b(T) & = &  b_3\left(\frac{T_0}{T} \right)^3\,,
\quad
\label{ew:ab:T}
\eeqn
where $T_0$ is the critical temperature in the pure gauge case:
$T_0 \equiv T_{SU(3)} = 270\, \mbox{MeV}$.

\begin{figure}[!thb]
\begin{center}
\includegraphics[width=85mm,clip=true]{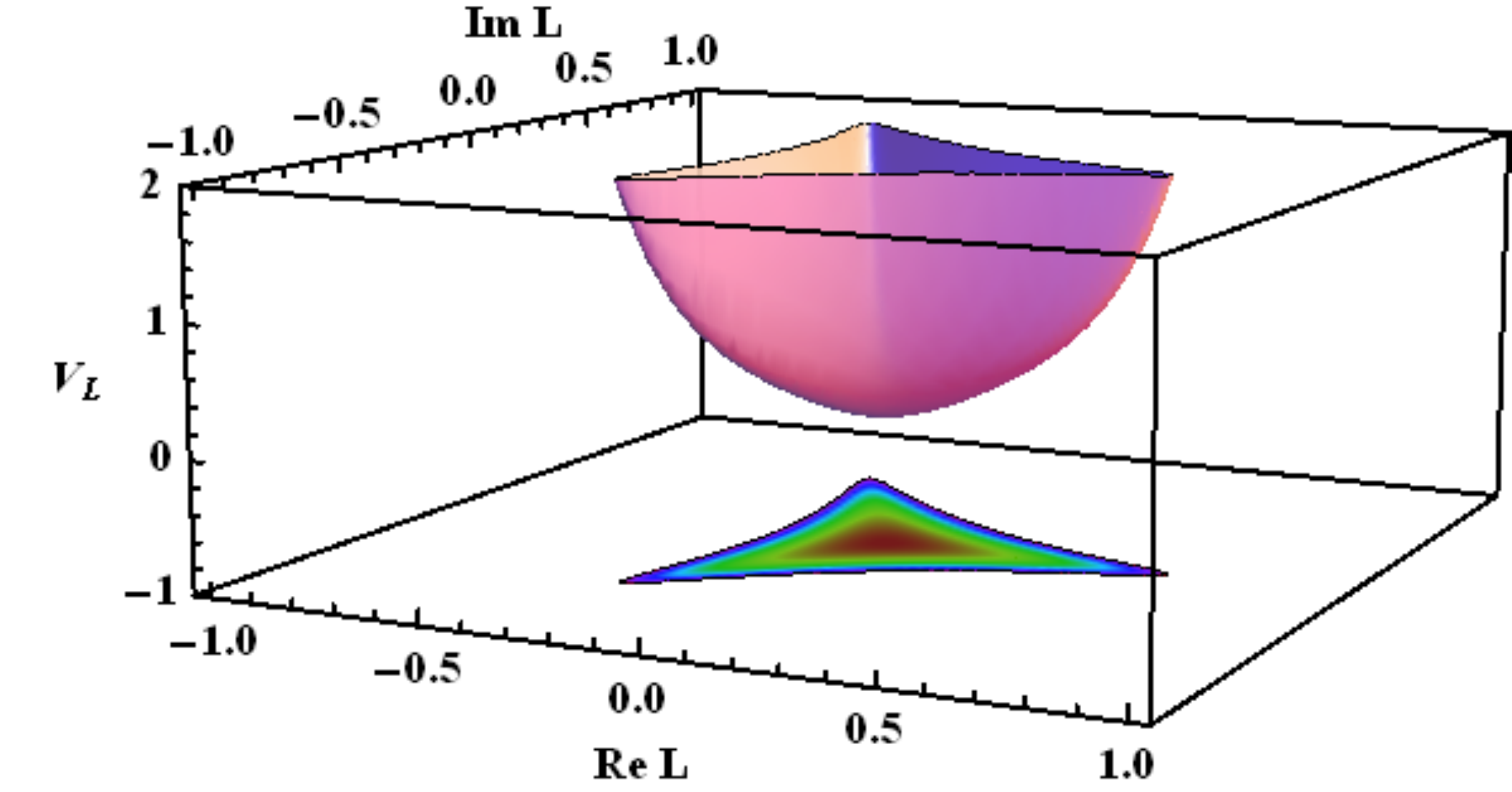}\\[-25mm]
\hskip -40mm $\mathbf{T=0.8 \, T_0}$\\[27mm]
\includegraphics[width=85mm,clip=true]{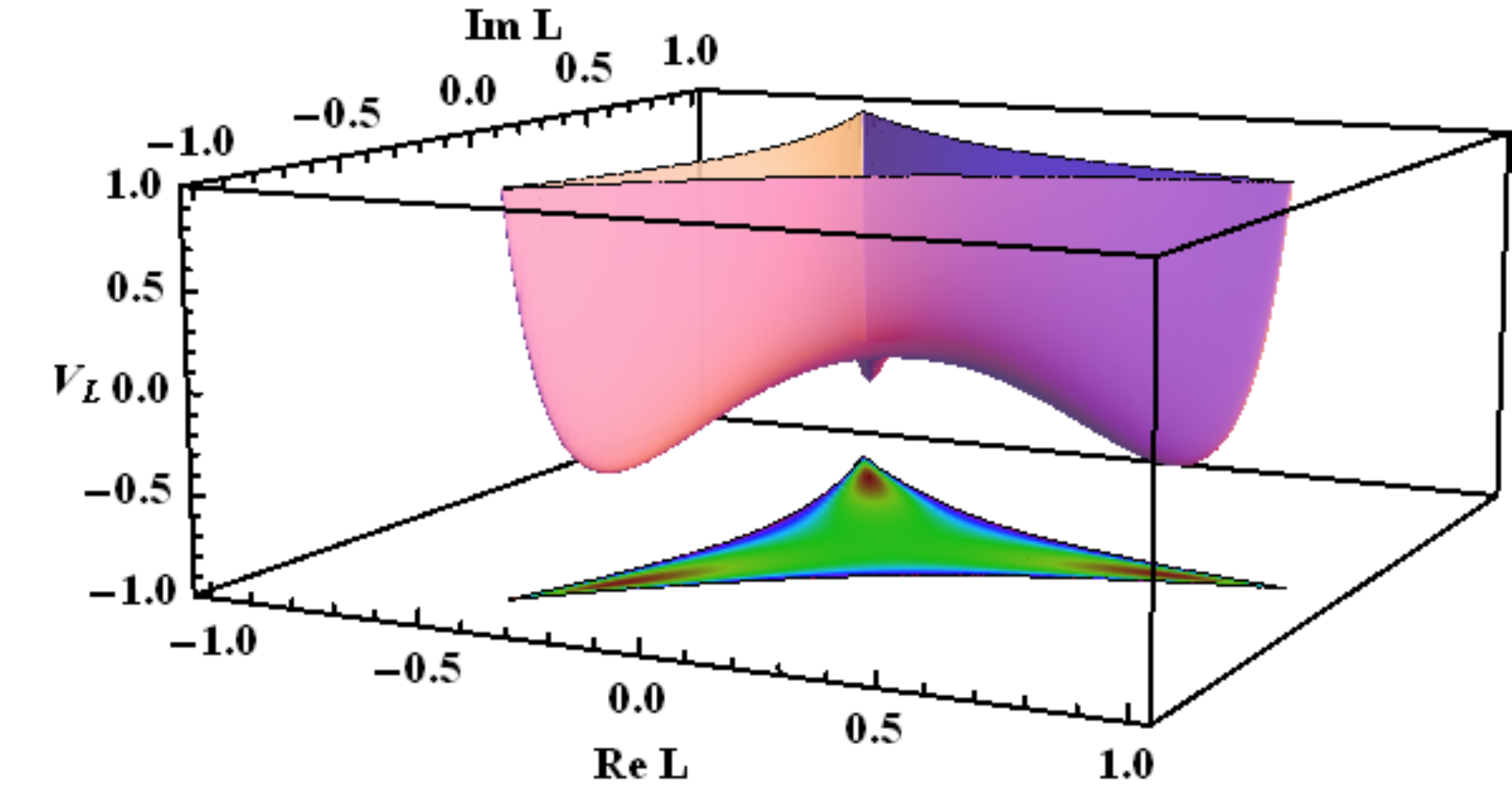}\\[-23mm]
\hskip -40mm $\mathbf{T=1.2 \, T_0}$\\[20mm]
\end{center}
\caption{The phenomenological potential of the Polyakov loop~\eq{eq:V} at
$T=0.8 \, T_0$ (top) and $T=1.2 \, T_0$ (bottom).}
\label{fig:VL:potential}
\end{figure}

The phenomenological parameters in Eq.~\eq{ew:ab:T} are
\beqn
\label{eq:V:parameters}
\begin{array}{llllll}
a_0 & = & 16\,\pi^2/45 \approx 3.51\,, & \qquad a_1 & = & -2.47\,, \\
a_2 & = & 15.2\,,                            & \qquad b_3 & = & -1.75\,.
\end{array}
\eeqn
Consistently with the original definition~\eq{eq:L}, the potential~\eq{eq:V} limits the
values of the Polyakov loop to the interval $L^* L \leqslant 1$ because of a logarithmic
divergence in \eq{eq:V}. The value $L^* L = 1$ is reached in the limit $T\rightarrow\infty$.
In the confined phase, $T < T_0$, the potential has one trivial minimum
[as illustrated in Figure~\ref{fig:VL:potential} (top)], whereas in the deconfined phase,
$T > T_0$, the potential has three degenerate minima [see Fig.~\ref{fig:VL:potential}
(bottom)]. The latter are mutually connected by $\Z_3$ transformations.

The set of parameters~\eq{eq:V:parameters} is obtained by demanding the following
requirements valid for pure $SU(3)$ gauge theory~\cite{ref:ratti08}:
\begin{itemize}
\item[(i)] the Stefan-Boltzmann limit is reached at $T\to \infty$;
\item[(ii)] a first-order phase transition takes place at $T=T_0$;
\item[(iii)] the potential describes well lattice data
for the Polyakov loop and for the thermodynamic functions such as
pressure, energy density and entropy.
\end{itemize}

The uncertainties for the parameters in Eq.~\eq{eq:V:parameters} are:
$6\,\%$ for $a_1$, less than $3\,\%$ for $a_2$, and about $2\,\%$ for
$b_3$~\cite{ref:ratti08}.

\subsection{Total free energy and observables}

The free energy of the system is given by
\beqn
\Omega = - \frac{T}{V_{3d}} \ln \cZ\,,
\label{eq:Omega:formal}
\eeqn
where $V_{3d}$ is the spatial volume and $\cZ$ is the partition function
\beqn
\cZ = \int \dD \psi \dD \bar\psi \dD \sigma \dD \vec \pi
\exp\Bigl[- i \int\limits_0^{1/T} d t \int d^3 x \, {\cal L} \Bigr]\,.
\label{eq:Z:formal}
\eeqn
In the mean field approximation, treating the meson fields classically and keeping
only the sigma direction in the mesonic sector, the free energy is given by
\beqn
\Omega(\sigma,L;T,B) & = & V_\phi(\sigma,\vec\pi) + V_{L}(L,T)
+ \Omega_q(\sigma,L,T) \,, \nonumber\\
\label{eq:Omega}
\eeqn
where the potential $V_\phi$ for the chiral field is given in Eq.~\eq{eq:V:sigma}
and the potential for the Polyakov loop $V_{L}$ is defined by Eq.~\eq{eq:V}.
In \eq{eq:Omega} we explicitly show the dependence of the free energy on the external
magnetic field $B$, the temperature $T$, and the mean field value of the chiral field
$\sigma$, as well as the expectation value of the Polyakov loop $L$.

The temperature and the magnetic field impose the fixed external conditions whereas
the expectation values of the dynamical fields are obtained by minimization of the free
energy~\eq{eq:Omega} with respect to the dynamical variables:
\beqn
\frac{\partial \Omega}{\partial X} = 0 \,, \qquad X = \sigma,\ L\,.
\label{eq:Omega:min}
\eeqn

The mass of the $\sigma$-mesons is given by the
curvature of $\Omega$ evaluated at the global minimum~\eq{eq:Omega:min}
of the free energy, i.e.:
\beqn
m_\sigma^2(T,B)  = \frac{\partial^2 \Omega}{\partial \, \sigma^2}\,.
\label{eq:masses}
\eeqn
Notice that the mass~\eq{eq:masses} depends implicitly on the temperature $T$ 
and the strength of the magnetic field $B$.  The zero-temperature and zero-field value 
of the $\sigma$-meson mass (defined in the preceding Section) is related to
Eq.~\eq{eq:masses} as follows
 $m_\sigma \equiv m_\sigma(T=0,B=0)$.

\section{Free energy at one loop}
\label{sec:free-energy}

We compute the free energy of the system in the one-loop approximation. To this end,
we functionally integrate over the quark fields and take the integral over the fluctuations
of the mesonic degrees of freedom near the minimum of the free energy, determined by
Eq.~\eq{eq:Omega:min}. These integrals yield thermal determinants, either in the background
of the magnetic field $B$ or for vanishing field.

The one-loop correction coming from quarks can be written in the form:
\beqn
\exp\Bigl[i \frac{V_{3d}}{T} \, (\Omega_u + \Omega_d)\Bigr] & {=} &
\int \dD \bar \psi \dD \psi \, \exp\Bigl[ - i \int d^4 x\, {\cal L}_q \Bigr]
\nonumber\\
& & \hskip -25mm = \prod_{q=u,d} \frac{\det_{T} \left[i \Dirac^{(q)} - g(\sigma
+i\gamma _{5} \vec{\tau} \cdot \vec{\pi} )\right]}{
\det \left[i \dirac - g(\sigma +i\gamma _{5} \vec{\tau} \cdot \vec{\pi} )\right]}
\,,\quad
\label{eq:Omega:quark}
\eeqn
where the quark Lagrangian ${\cal L}_q$ is given in Eq.~\eq{eq:L:psi}. The notation
``${\det}_T$'' means that the determinant is taken at nonzero temperature, while
``${\det}$'' indicates that we consider the $T=0$ case.

The numerator in Eq.~\eq{eq:Omega:quark} involves the covariant derivative~\eq{eq:D}
because we calculate the contribution of the fermion loop in the presence of an external
magnetic field and in the background of the nontrivial Polyakov loop.  The magnetic
field $B$ and the Polyakov loop $L$ affect the chiral degrees of freedom via the covariant
derivative. Thus, the integration over quark fields provides an effective potential for the
meson fields, $\sigma$ and $\vec \pi$, and the Polyakov loop, $L$. The coefficients of this
potential should in general depend on the strength of the magnetic field $B$ and on the
temperature $T$. The determinant in the numerator of Eq.~\eq{eq:Omega:quark} is regularized
by a similar determinant in the denominator which is calculated at $T=0$, $B=0$ and in the
absence of the confining background. Thus, the quark one-loop correction is expressed in terms
of a ratio of fermionic determinants. In what follows we assume
that the vacuum of the model is defined by $\langle\sigma\rangle = f_{\pi}$, $\langle\vec\pi\rangle =0$.

Following Ref.~\cite{Agasian:2008tb} it is convenient to regroup in Eq.~\eq{eq:Omega:quark}
the ratio of the determinants, and represent the free energy of each species of the light
quarks as follows:
\beqn
e^{i V_{3d} \, \Omega_q/T}  & = &
\left[\frac{\det(i \Dirac^{(q)} - m_q )}{\det (i \dirac - m_q)} \right]  \nonumber\\
& & \cdot
\left[\frac{\det_T(i \Dirac^{(q)} - m_q)}{\det (i \Dirac^{(q)} - m_q )}\right]\,,
\label{eq:Omega:quark2}
\eeqn
where $q$ stands for either $u$ or $d$. The first ratio corresponds to the response
of the quark loops to the external fields at zero temperature. This term represents
the vacuum contribution, $\Omega^{\mathrm{vac}}_{q}$. The second ratio in
Eq.~\eq{eq:Omega:quark2} represents the finite-temperature correction due to thermal
fermion excitations. As in Ref.~\cite{Agasian:2008tb}, we call this contribution the
paramagnetic part of free energy, $\Omega^{\mathrm{para}}_{q}$.

In this paper, as mentioned previously, we consider two approaches for the PLSMq, one that includes
$B$-dependent vacuum effects and one that does not. We intentionally ignore pure vacuum corrections,
in order to compare our results with what is found in the absence of a magnetic background in the
literature of the linear sigma model with quarks, coupled or not to Polyakov loops (e.g.,
Ref. \cite{Scavenius:2000qd}). However, since this is a theory with spontaneous symmetry breaking
(besides an explicit breaking), a nonzero condensate modifies the quark mass, and vacuum subtractions
are more subtle. At the end, there are finite logarithmic contributions that survive renormalization, sometimes
called zero-point corrections, but that are usually discarded phenomenologically in the LSMq. These
contributions have been studied by the authors of Ref. \cite{Mocsy:2004ab} in the LSMq.
Vacuum contributions were considered at finite density in the perturbative massive Yukawa model with
exact analytic results up to two loops in Ref. \cite{Palhares:2008yq,thesis} and, more specifically, in optimized
perturbation theory at finite temperature and chemical potential in Ref. \cite{Fraga:2009pi},
also comparing to mean-field theory. More recently, this issue was discussed in a comparison
with the Nambu-Jona--Lasinio model \cite{Boomsma:2009eh}. A very recent careful and detailed study
in the (Polyakov loop extended) quark-meson model, with special attention to the chiral limit, can be found in
Refs. \cite{Skokov:2010wb}.

Thus, the one-loop correction from quarks to the free energy of the system is given by
\beqn
\Omega_{q}(B,T) = \sum_{f=u,d} \Bigl[\Omega^{\mathrm{vac}}_{q_f}(B) +
\Omega^{\mathrm{para}}_{q_f}(B,T)\Bigr]\,. \
\label{eq:Vqsum}
\eeqn
Below we discuss each term in this equation separately.

\subsection{Vacuum contribution, $\Omega^{\mathrm{vac}}_q$}
\label{sec:vacuum}

The vacuum contribution can be expressed as the Heisenberg--Euler energy density\footnote{In Ref.~\cite{ref:review:Dunne} one 
can find results for spinor and scalar determinants in the presence of external fields.}:
\beqn
\Omega^{\mathrm{vac}}_{q}(B) & = &
\frac{1}{i V_{4d}} \log \left[\frac{\det(i \Dirac^{(q)} - m_q )}
{\det (i \dirac - m_q)} \right]
\label{eq:Omega:q:vac}\\
& & \hskip -15mm
= N_c \cdot \frac{(q B)^2}{8 \pi^2} \! \int\limits_0^\infty \!\frac{d s}{s^3}
\left(\frac{s}{\tanh s} - 1 - \frac{s^2}{3}\right) \, e^{- s \, m^2_q/(q B)}\,, \
\nonumber
\eeqn
where we have already accounted for the color degeneracy of the quark degrees of
freedom.

Since the covariant derivative $\Dirac^{(q)}$ contains not only the electromagnetic
but also the non-Abelian gluon fields, one could expect that non-Abelian features
should also appear in Eq.~\eq{eq:Omega:q:vac}. Notice, however, that the Polyakov
loop contribution is trivial since, for $T=0$, it can be put away by a simple shift in the
integral over the zeroth component of the momentum. So, for the vacuum contribution
from quarks we can ignore its presence in the covariant derivative. On the other hand,
as we discuss below, at non-zero temperature the presence of the gluon field leads
effectively to a shift of the Matsubara frequencies, thus contributing to the paramagnetic
(second) part of~\eq{eq:Vqsum}.

The expression above corresponds, of course, to the quark vacuum correction in
the presence of the external magnetic field
\beqn
\Omega_{q}^{(B)}&=&-\frac{N_{c}}{\pi}
\sum_{f=u,d}|q_{f}|B \left[ \left(\sum_{n} I_{B}^{(1)}(M_{nf}^{2})\right)
\right. \nonumber \\
&& - \left.
\frac{I_{B}^{(1)}(m_{f})}{2}\right] \,
\eeqn
minus the vacuum correction in the absence of the field
\beqn
\Omega_{q}^{(0)}&=& 2N_{c}\sum_{f=u,d}I_{B}^{(3)}(m_{f}^{2}) \, ,
\eeqn
where we have defined the integral
\beqn
I_{B}^{(d)}(M^{2})&=& \int \frac{d^{d}p}{(2\pi)^{d}} \sqrt{p^{2}+M^{2}}
\eeqn
and the quark mass in the presence of $B$
\beqn
M_{nf}^{2}&=& m_{f}^{2}+2n|q_{f}|B \, .
\eeqn
One can control the divergences in the integrals above via dimensional regularization
and compute the difference $\Omega_{q}^{(B)}-\Omega_{q}^{(0)}$ in the $\overline{\rm MS}$
scheme, obtaining
\beqn
\Omega^{\mathrm{vac}}_{q}(B) & = &
-\frac{N_{c}}{2\pi^{2}}\sum_{f=u,d}(q_{f}B)^{2} \left[ \zeta'(-1,x_{f})
\right.\nonumber \\
&-&\left.
\frac{1}{2}(x_{f}^{2}-x_{f})\log x_{f} +\frac{x_{f}^{2}}{4}
\right.\nonumber \\
&+&\left.
\frac{1}{12} \log\left(\frac{\Lambda^{2}}{2|q_{f}|B}  \right) \right] \, ,
\label{eq:vacuum}
\eeqn
where $x_{f}\equiv m_{f}^{2}/(2|q_{f}|B)$, $\zeta'$ is the derivative with respect to
the first argument of the Riemann-Hurwitz $\zeta$ function \cite{AS}, and $\Lambda$ is the
renormalization subtraction scale. This expression is in agreement with the results
of Refs. \cite{Ebert:1999ht,Menezes:2008qt}, except for the scale-dependent term.
Since the scale-dependent term
is independent of the meson fields and we are concerned only with the effective potential
of the theory, we can discard it. Here our normalization is such that this correction vanishes
in the limit of $B=0$. One arrives at the same result using the Heisenberg-Euler approach
and imposing the normalization mentioned above.

\begin{figure}[!thb]
\begin{center}
\includegraphics[width=80mm,clip=true]{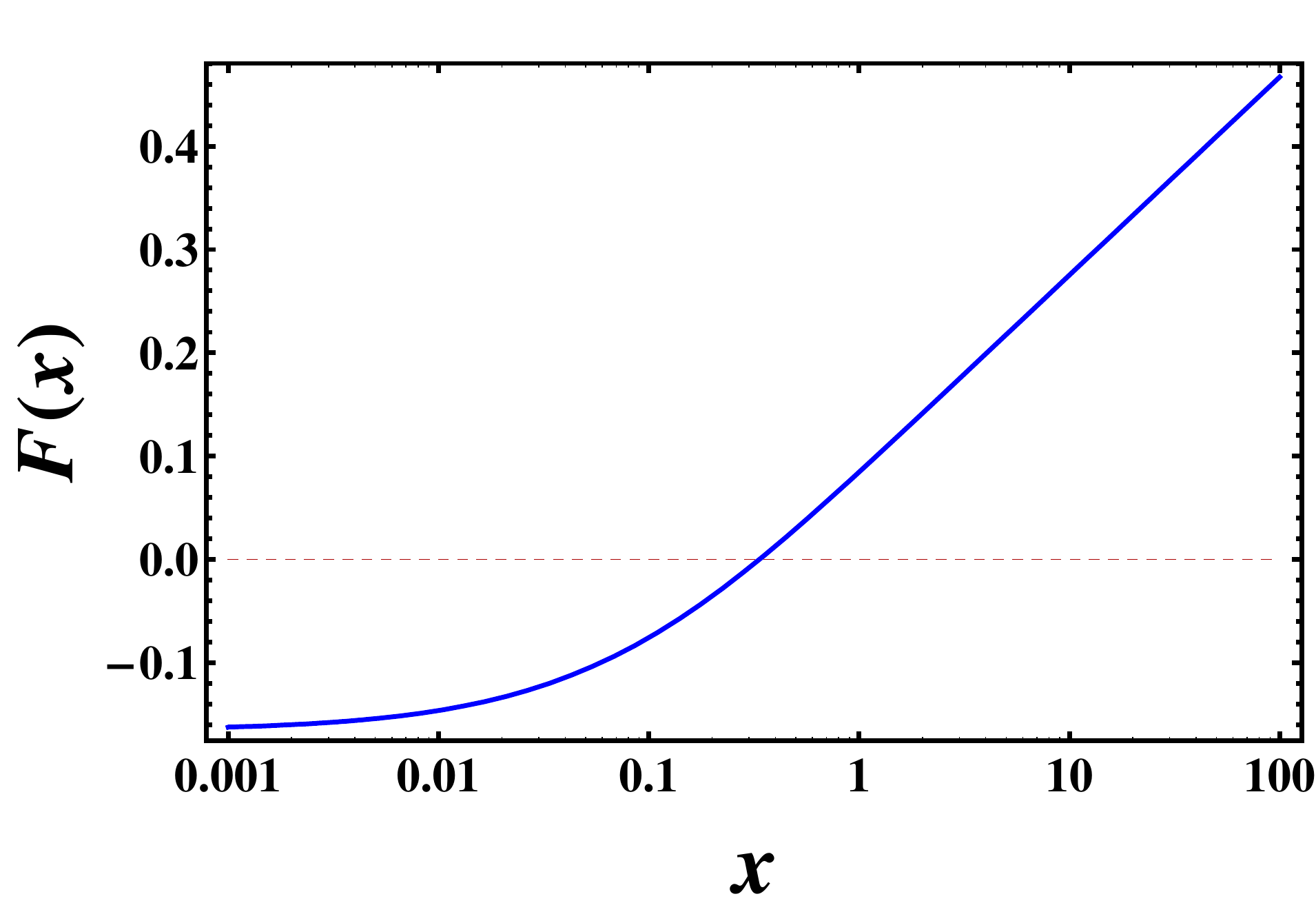}
\end{center}
\vskip -5mm
\caption{
The function $F(x)$, Eq.~\eq{eq:F}. }
\label{fig:funcF}
\vskip -3mm
\end{figure}

We can cast the correction to the effective potential in a dimensionless form normalizing every
quantity by a mass scale. For this purpose, we choose $v$ as our mass unit and define:
\beqn
\xi&\equiv& \frac{\sigma}{v}\,, \qquad
b\equiv \frac{eB}{v^{2}}\,, \qquad
t\equiv \frac{T}{v}\,.
\eeqn
It is also convenient to write the quark electric charge as $q_{f}=r_{f}eB~{\mathrm{sign}}(q_{f})$, so
that $r_{u}=2/3$ and $r_{d}=1/3$. Then, the vacuum
one-loop contribution to the effective potential is given by
\beqn
\frac{V_{\rm vac}(\xi,b)}{v^{4}}&=& -\frac{N_{c}b^{2}}{2\pi^{2}} \sum_{f=u,d}r_{f}^{2}
F\left( \frac{g^{2}\xi^{2}}{2r_{f}b} \right),
\eeqn
where the function $F$ is defined as
\beqn
F(x)&\equiv&
\zeta'(-1,x_{f}) -
\frac{1}{2}(x_{f}^{2}-x_{f})\log x_{f} +\frac{x_{f}^{2}}{4} \, ,
\label{eq:F}
\eeqn
and displayed in Figure~\ref{fig:funcF}. For small values of its argument it behaves as
\beqn
F(x) &=& -0.165421 + \frac{1}{2} x \log x + {\cal O}(x)\, ,
\eeqn
whereas for large a argument one finds
\beqn
F(x) &=& \frac{1}{12} (1 + \log x) + {\cal O}(x^{-2}) \,.
\eeqn

Since the constituent quark mass $m_q$ is linear in $\sigma$, the external magnetic field
modifies the classical potential~\eq{eq:V:sigma} for the $\sigma$ field. One can see from
the behavior of this correction that its effect is to increase the value of the condensate
and deepen the absolute minimum of the effective potential, as expected from previous
results on magnetic catalysis.

Finally, for the sake of completeness, we discuss the weak and strong field limits of
the $T=0$ quark free energy~\eq{eq:Omega:q:vac}. The leading term in the weak-field limit
($|q B| \ll m_q^2$) is
\beqn
\Omega^{\mathrm{vac}}_{q}(B) & = & - \frac{N_c}{360 \pi^2} {\left(\frac{q B}{m_q}\right)}^4 \left\{1 +
O\left[\left(\frac{q B}{m_q^2}\right)^2\right] \right\}.\quad
\eeqn

The strong-field limit ($|q B| \gg m_q^2$) with on-shell renormalization ($ \Lambda = m$)
is given by~\cite{ref:review:Dunne}:
\beqn
\Omega^{\mathrm{vac}}_{q}(B) & = & - \frac{N_c}{24 \pi^2} \, (q B)^2 \,
\ln \left[C_2\,\frac{2 e B}{m_q^2}\right]
\nonumber\\
& & - \frac{N_c}{8 \pi^2} \, q B \, m_q^2 \, \ln \left[C_1\,\frac{2 e B}{m_q^2}\right]
\label{eq:Omega:q:strong}\\
& & - \frac{N_c}{16 \pi^2} \, m_q^4 \, \ln \left[C_0\,\frac{2 e B}{m_q^2}\right]
+ O\left(\frac{m_q^6}{q B}\right)\,.
\nonumber
\eeqn
where
$C_0 = e^{3/2 - \gamma} = 2.51629$, $C_1 = {\mathrm{e}}/2 \pi = 0.432628$, $C_2 =
\mathcal{A}^{-12} = 0.0505368$,
$\gamma = 0.577216$ is the Euler constant, and $\mathcal{A} = 1.28243$ is the Glaisher
constant\footnote{In Ref.~\cite{Fraga:2008qn}, the quark vacuum term was not included, the
charged pions  vacuum contribution being considered instead. Both treatments increase
the value of the condensate and deepen the effective potential. In this paper we choose
to treat the mesonic fields classically.}.

Numerical tests show that only in the limit of very high fields, in units of $v$,
the vacuum corrections bring a non-negligible modification to the effective potential.

\subsection{Paramagnetic contribution, $\Omega^{\mathrm{para}}_q$}

The paramagnetic temperature-induced contribution comes from the second determinant
ratio in Eq.~\eq{eq:Omega:quark2}. In general, a quark determinant can be formally written
as the product over the eigenvalues $\Lambda_{n}$ of the corresponding Dirac operator, i.e.:
\beqn
\det [i\Dirac + m_q]_T \equiv \prod_{n} \frac{\Lambda_{n}}{T}\,,
\label{eq:det1}
\eeqn
where $\Lambda_{n}$ satisfy the eigenvalue equation
\beqn
[i\Dirac + m_q] \psi_n = \Lambda_n \psi_n\,.
\eeqn
It is important to stress that in Eq.~\eq{eq:det1} we omitted a $T$-independent constant
which would anyway be subtracted at the end of the calculation.

The algebraic equation for eigenvalues $\Lambda_n$ reads as follows:
\beqn
& & (p_0 - A^{ii}_0)^2 - p^2_z - (m_q - \Lambda)^2  \nonumber \\ 
& & \hskip 30mm - (2 n + 1 - 2 s) |q| B = 0\,, 
\label{eq:Lambda:1}
\eeqn
where $q$ is the electric charge of the quark, $n =0,1,2,\dots$ is the quantum number that
labels the Landau levels and $s = \pm 1/2$ is the projection of the spin of the quark eigenstate
onto the $z$-axis. Here $A^{ii}_0$ is the $i$th diagonal component of the $SU(3)$ gauge
field. We assume no sum over color indices $i=1,2,3$ unless explicitly indicated.
From \eq{eq:Lambda:1} one finds
\beqn
\Lambda^{(i)}_{\pm} {=} m_q \pm {\bigl[ (p_0 - A^{ii}_0)^2 - p^2_z -
(2 n + 1 - 2 s) |q| B \bigr]}^{1/2}\!\!. \qquad
\label{eq:Lambda:2}
\eeqn
At zero magnetic field the contribution to the effective potential coming from~\eq{eq:det1}
can be written as follows:
\beqn
\frac{T}{V_{3d}} \ln \det [i\Dirac + m]_T = T\sum_{\ell \in \Z} \int \frac{d^3 p}{(2\pi)^3}
\ln\Bigl[\frac{\Lambda^{(i)}_+ \Lambda^{(i)}_-}{T^2}\Bigr]\,, \qquad 
\label{eq:V:psi}
\eeqn
where $\ell$ labels the Matsubara frequencies $\omega_\ell$.
The zeroth component of the momentum is related to the Matsubara frequency as follows:
\beqn
p_0 = - i p_4 \equiv - i \omega_\ell, \qquad \omega_\ell = 2 \pi T (\ell + 1/2)\,,\qquad 
\eeqn
where the odd frequency takes into account the anti-periodicity of the quark fields
along the temporal direction.

In the presence of a magnetic field $B$, the integral over the phase space
in Eq.~\eq{eq:V:psi} is modified to
\beqn
\int \frac{d^3 p}{(2\pi)^3}
\mapsto \frac{|q|B}{2\pi} \sum_{n=0}^\infty \int \frac{d p_z}{2\pi}\,,
\label{eq:mapping}
\eeqn
where the integer $n$ labels the Landau levels.

Using the mapping \eq{eq:mapping}, the explicit expression for the fermion determinant
\eq{eq:V:psi} and the eigenvalues~\eq{eq:Lambda:2}, we can rewrite the quark-induced
potential~\eq{eq:Omega:quark2} as follows:
\beqn
\Omega^{\mathrm{para}}_q(\sigma,\Phi) & = &
\frac{T}{i V_{3d}} \ln \left[\frac{\det_T(i \Dirac^{(q)} - m_q)}{\det (i \Dirac^{(q)} - m_q )}\right]
\nonumber\\
& = &  - \frac{|q| B T}{2\pi} \sum_{i=1}^3 \sum_{s = \pm \frac{1}{2}}
\sum_{n=0}^\infty \sum_{\ell\in\Z} \quad
\label{eq:Vpsi:1}\nonumber \\
& & \int\limits_{-\infty}^{+\infty}
\frac{d p_z}{2 \pi} \ln \Bigl[\Bigl(\frac{\omega_\ell}{T}
+ \frac{A^{ii}_4}{T}\Bigr)^2 + \frac{\omega^2_{sn}(p_z,\sigma)}{T^2} \Bigr]\,,
\nonumber \\
\eeqn
where the dispersion relation for quarks in the external magnetic field is
\beqn
\omega_{sn}(p_z,\sigma) = {\bigl[m_q^2(\sigma) + p_z^2 + (2 n + 1 - 2 s) |q| B\bigr]}^{1/2}\,,
\label{eq:omega:sn}
\eeqn
with constituent quark mass $m_q(\sigma)= g \sigma$.

With a suitable identification, $\gamma = A^{ii}_4/T$ and $\theta = \omega_{sn}/T$,
the sum over Matsubara frequencies in Eq.~\eq{eq:Vpsi:1} can be done explicitly
with the help of the following formula (with arbitrary real $\gamma$ and $\theta$):
\beqn
& & \hskip -7mm \sum_{\ell \in \Z} \ln[(2 \pi \ell + \pi + \gamma)^2 + \theta^2] =
4 \sum_{j=1}^{\infty} \ln(2 \pi j)
\label{eq:sum}\\
& & + \theta + \ln (1 + e^{i \gamma} e^{- \theta}) + \ln (1 + e^{ - i \gamma} e^{- \theta})\,.
\nonumber
\eeqn

The first term in the right hand side of equation above is divergent. In the language of the effective
potential~\eq{eq:Vpsi:1}, the divergent term corresponds to a contribution from massless
fermions plus an additive constant. We ignore this term since it depends neither on chiral
nor on confining variables. The second term leads to a $T$--independent contribution
to \eq{eq:Vpsi:1}, which is already taken into account by the vacuum energy~\eq{eq:Omega:q:vac}.
The last two terms lead to the regularized paramagnetic free energy for the chiral and confining fields:
\beqn
\Omega^{\mathrm{para}}_q = - \frac{|q| B T}{2 \pi} \sum_{s = \pm \frac{1}{2}} \sum_{n=0}^\infty
\int\limits_{-\infty}^{+\infty} \frac{d p_z}{2 \pi} \, W_T[\omega_{sn}(p_z,\sigma),\Phi]\,, \quad
\label{eq:V2}
\eeqn
where
\beqn
W_T(\omega,\Phi) & = & {\mathrm{Tr}}_c\, \Bigl[\ln \Bigl(\bbbone +
\Phi \, e^{- \omega/T} \Bigr) + c.c. \Bigr]
\label{eq:W:T}\\
& = & \sum_{i=1}^3 \ln \Bigl(1 + 2 e^{- \omega/T} \cos \varphi_i  +
e^{- 2 \omega/T}\Bigr)\,.
\nonumber
\eeqn
Here the trace Tr${}_c$ is taken in the color space and the angles $\varphi_i$ are defined in
the diagonal representation~\eq{eq:Phi} of the untraced Polyakov loop~$\Phi$ [the Polyakov
loop is defined in Eq.~\eq{eq:L}]. 
Notice that the parameters $\varphi_i$ -- which are the phases of the diagonal components 
of the untraced Polyakov loop -- enter the partition function exactly like imaginary chemical potentials. 

The paramagnetic contribution, given by Eqs.~\eq{eq:V2} and \eq{eq:W:T}, can be simplified
further by expanding the logarithmic function in Eq.~\eq{eq:W:T} in the following series:
\beqn
& & \ln \Bigl(1 + 2 e^{- \omega/T} \cos \varphi  + e^{- 2 \omega/T}\Bigr)
\nonumber \\
& & \qquad = - 2 \sum_{k=1}^{\infty} \frac{(-1)^k }{k}
\cos(k \varphi)\, e^{- k \omega/T}\,.
\label{eq:ln:series}
\eeqn
Next, we use the simple relation
\beqn
\int\limits_{-\infty}^{+\infty} d p_z \, \exp\left\{- \frac{\sqrt{p_z^2 +
\mu^2}}{T} \right\} = 2\mu \,K_1\left(\frac{\mu}{T}\right)\,,
\eeqn
where $K_l(x)$ is the modified Bessel function of the second kind (the MacDonald function) and
order $l$ \cite{AS}. Finally, we express the free energy~\eq{eq:V2} in terms of the sums only:
\beqn
\Omega^{\mathrm{para}}_q & = & \frac{|q| B T}{\pi^2} \sum_{s = \pm \frac{1}{2}} \sum_{n=0}^\infty \sum_{k=1}^\infty
\frac{(-1)^k}{k} \mathrm{Re}\, \left[\mathrm{Tr} \Phi^k \right] \nonumber\\
& & \times \mu_{sn}(\sigma)\,  K_1\left[\frac{k}{T} \mu_{sn}(\sigma)\right] \, ,
\label{eq:para}
\eeqn
where $\mu_{sn} \equiv \omega_{sn}(p_z=0,\sigma)$ is the energy of the $n$th Landau
level~\eq{eq:omega:sn} at zero longitudinal momentum
\beqn
\mu_{sn}(\sigma)  = {\bigl[g^2\sigma^2 + (2 n + 1 - 2 s) |q| B\bigr]}^{1/2}\,,
\label{eq:mu:sn}
\eeqn
and the untraced Polyakov loop is defined in Eq.~\eq{eq:Phi}, so that
\beqn
\mathrm{Re}\, \left[\mathrm{Tr} \Phi^k \right] = \sum_{i=1}^3 \cos(k \varphi_i)\,.
\eeqn
The integer number $k$ corresponds to the winding number of the Polyakov loops.

\begin{figure}[!thb]
\begin{center}
\includegraphics[width=85mm,clip=true]{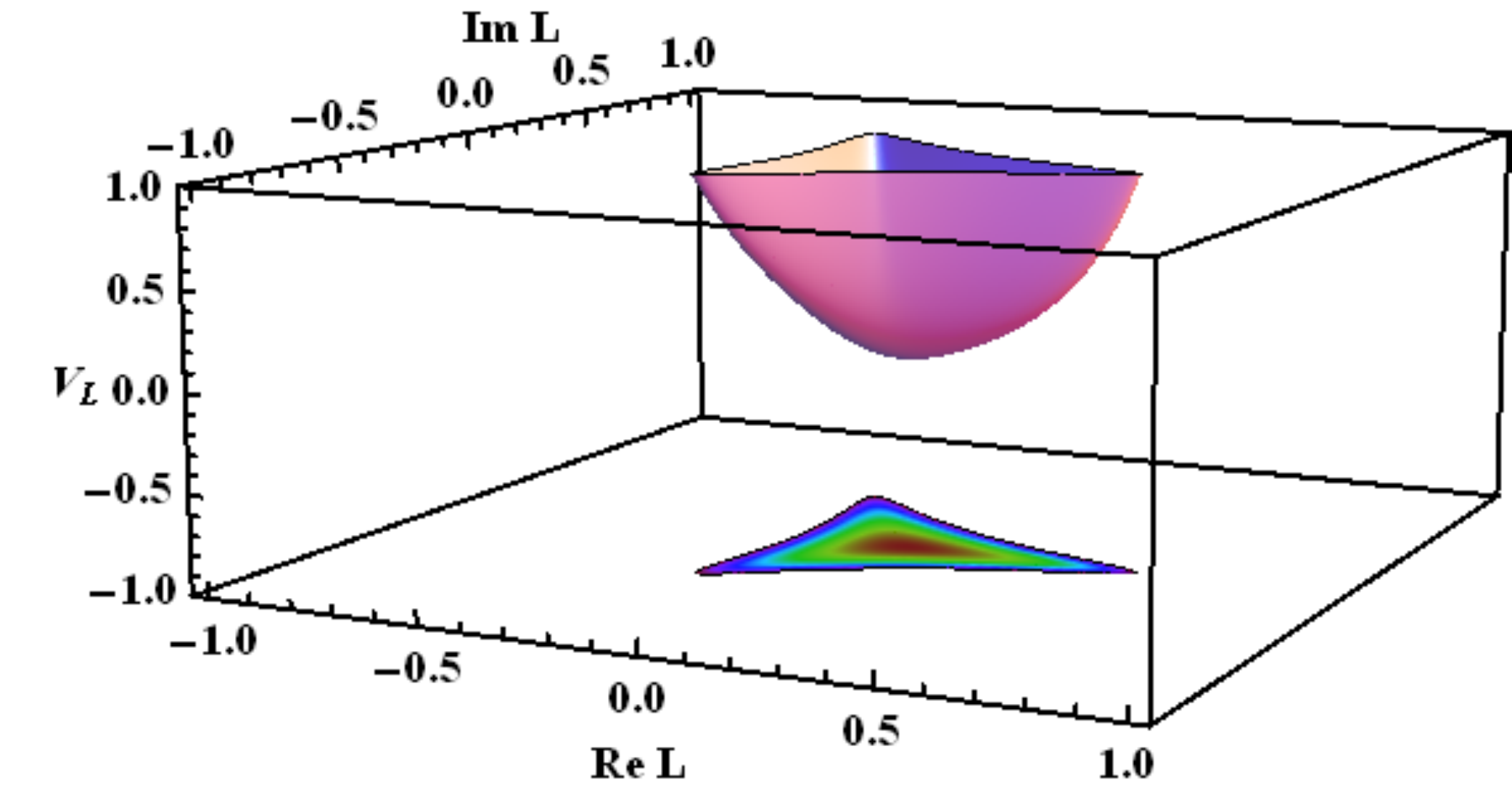}\\[-25mm]
\hskip -40mm $\mathbf{T=0.8 \, T_0}$\\[27mm]
\includegraphics[width=85mm,clip=true]{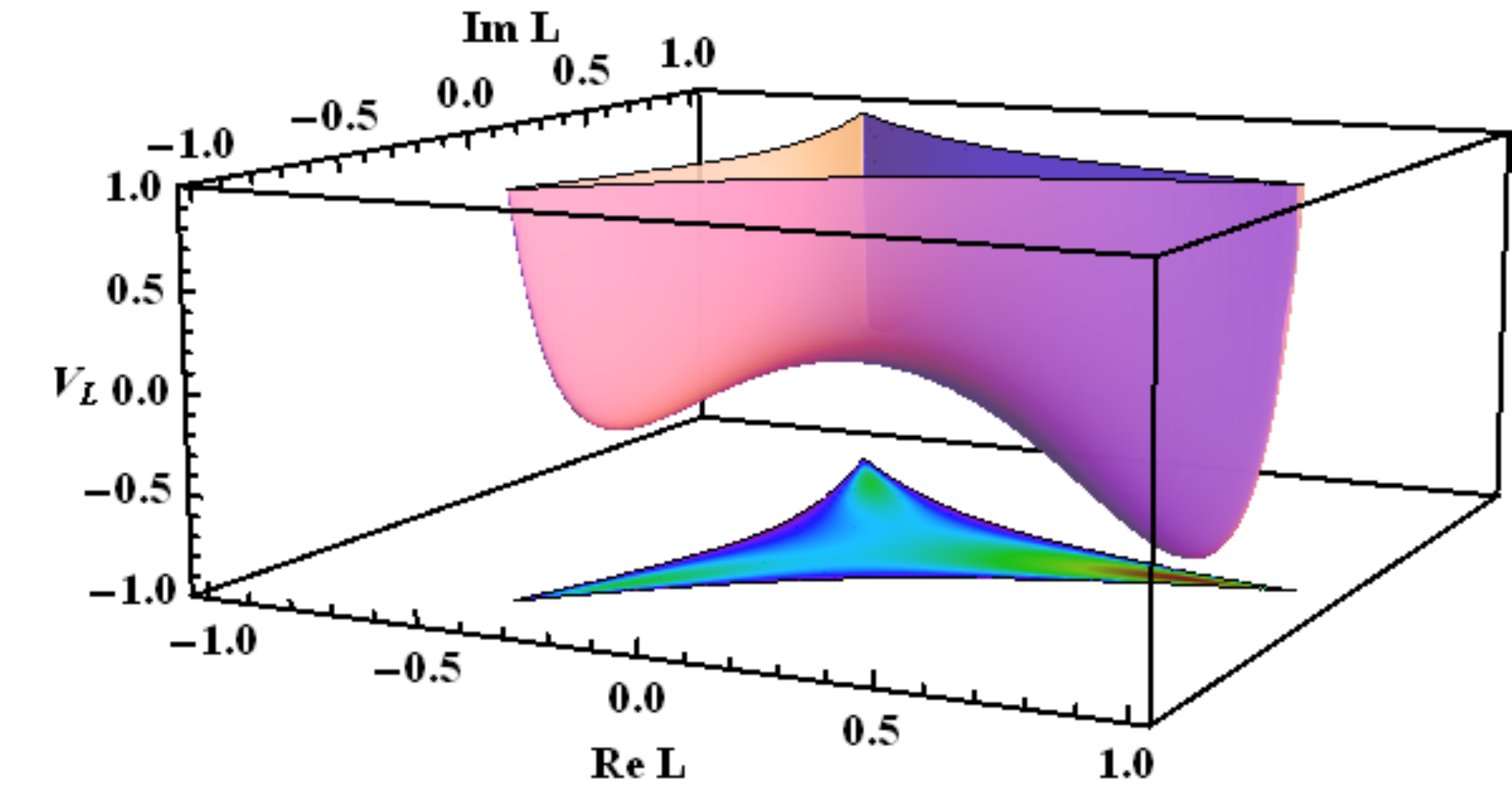}\\[-23mm]
\hskip -40mm $\mathbf{T=1.2 \, T_0}$\\[20mm]
\end{center}
\caption{The same as in Fig.~\ref{fig:VL:potential} but with the paramagnetic
contribution~\eq{eq:para} included. The potential is shown at two temperatures:
$T=0.8 \, T_0$ (top) and $T=1.2 \, T_0$ (bottom). The magnetic field is given
by $\sqrt{e B} = 3 T$ in both cases.}
\label{fig:VL:potential:para}
\end{figure}

It is also convenient to express the paramagnetic contribution to the effective potential
in terms of dimensionless quantities as follows:
\beqn
\frac{V^{\rm para}(\xi,\phi_{1},\phi_{2},b,t)}{v^{4}}&=&
-\frac{bt^{2}}{2\pi^{2}} K(b/t^{2},\xi/t,\phi_{1},\phi_{2})\, ,
\label{eq:Vpara-dim}
\eeqn
where we defined the dimensionless function
\beqn
K(\xi,\phi_{1},\phi_{2},b,t) &=& \sum_{f=u,d}\sum_{s=\pm 1/2}\sum_{n=0}^{\infty}\sum_{i=1}^{3}
\nonumber \\
&& \hskip -20mm \times\,\int_{0}^{\infty} dx \log \left( 1 +  e^{-2\sqrt{x^{2}+\tilde\mu_{snf}(\xi,b)/t}}
\right.\nonumber \\ \nonumber \\
&& \hskip -15mm \left. + 2 e^{-\sqrt{x^{2}+\tilde\mu_{snf}(\xi,b)/t}} \cos\phi_{i}  \right)
\label{eq:funcK}
\eeqn
and the dimensionless version of Eq. (\ref{eq:mu:sn}), i.e.
\beqn
\tilde\mu_{snf}&=& \left[ g^{2}\xi^{2} + (2n+1-2s)r_{f}b  \right]^{1/2} \,.
\eeqn
%

\subsection{Paramagnetically-induced breaking of $\Z_3$}

One can see from Eq.~\eq{eq:para} that {\it the magnetic field drastically affects the
potential for the Polyakov loop}. Take, for example, the limit of a very strong field,
$\sqrt{|q| B} \gg m_q \equiv g \sigma$. Then, the leading contribution to Eq.~\eq{eq:para} is
given by the lowest Landau level, $n=0$, with spins oriented along the field, $s= + 1/2$,
at lowest winding number of the Polyakov loop, $k=1$. Thus, the leading term of the
strong-field expansion of the paramagnetic potential~\eq{eq:para} is
\beqn
\Omega^{\mathrm{para}}_q = - 3 \frac{g\sigma |q| B T }{\pi^2}
K_1\left(\frac{g\sigma}{T}\right) \mathrm{Re} \, L\,,
\label{eq:strong:field}
\eeqn
where the Polyakov loop variable is defined in Eq.~\eq{eq:L}. 
At fixed value of the magnetic field and temperature, the function~\eq{eq:strong:field} is
a nonmonotonic function of the field $\sigma$: it increases at small values of $\sigma$ 
and decreases at large values of $\sigma$. This fact will be essential for our discussion 
of the role of the vacuum contribution introduced in the previous subsection: the vacuum 
contribution makes the expectation value of the $\sigma$ field larger, eventually 
diminishing the role of the paramagnetic contribution~\eq{eq:strong:field} at large enough magnetic fields.

Notice that the paramagnetic contribution~\eq{eq:strong:field} is not
invariant under the $\Z_3$ symmetry, and therefore the paramagnetic
contribution~\eq{eq:strong:field} deforms the potential for the Polyakov loops. So, {\it it is
clear that the magnetic field tends to break the $\Z_3$ symmetry and induce deconfinement in
this model}.

In order to illustrate the effect of the paramagnetic term~\eq{eq:para} we plot in
Figure~\ref{fig:VL:potential:para} the potential for the Polyakov loop~\eq{eq:V} including the
correction induced by the paramagnetic interaction~\eq{eq:strong:field}. In the figure,
The magnetic field is $\sqrt{e B} = 3 T$.  We show the potential for two temperatures,
$T=0.8 \, T_0$ and $T=1.2 \, T_0$, in order to compare the plots with the pure gauge field
potential displayed in Figure~\ref{fig:VL:potential}. The effect of the magnetic field is clearly seen
in the deconfined phase: {\it the paramagnetic interaction induces the expectation value
of the Polyakov loop to be real-valued}.

\subsection{Complete effective potential}

It is, of course, also convenient to define dimensionless expressions for the classical
potential
\beqn
\frac{V_{cl}(\xi)}{v^{4}}&=&\frac{\lambda}{4}(\xi^{2}-1)^{2}-\tilde h \xi \,,
\label{eq:Vcl-dim}
\eeqn
where $\tilde h\equiv h/v^{3}$, and for and the potential for the Polyakov loops
\beqn
\frac{V_{P}(\phi_{1},\phi_{2},t,t_{0})}{v^{4}}&=& \frac{V_{L}(L,T)}{v^4}\,,
\label{eq:VP-dim}
\eeqn
where we used Eq. (\ref{eq:V}) and defined $t_{0}\equiv T_{0}/v$, besides expressing
$L$ and $L^{*}$ in terms of $\phi_{1}$ and $\phi_{2}$.

The total (dimensionless) effective potential can then be expressed as
\beqn
\frac{V_{eff}(\xi,\phi_{1},\phi_{2},b,t)}{v^{4}}&=&
\frac{V_{cl}(\xi)}{v^{4}}
+ \frac{V_{P}(\phi_{1},\phi_{2},t,t_{0})}{v^{4}} \nonumber\\
&+& \frac{V_{\rm vac}(\xi,b)}{v^{4}}
+ \frac{V^{\rm para}(\xi,\phi_{1},\phi_{2},b,t)}{v^{4}} \,,\nonumber\\
\label{eq:Veff-dim}
\eeqn
and it is this form of the effective potential that we use to determine the phase structure
of the \M\ effective field theory in the following section.

\section{Phase structure}
\label{sec:results}

In our numerical analysis we have considered the following situations:
\begin{itemize}
\item[(A)] The linear sigma model with thermal corrections, without the coupling to the Polyakov loop. This is a standard analysis and has been performed previously in \cite{quarks-chiral,ove,Scavenius:1999zc,Caldas:2000ic,Scavenius:2000qd,
Scavenius:2001bb,paech,Mocsy:2004ab,Aguiar:2003pp,Schaefer:2006ds,
Taketani:2006zg,Fraga:2004hp}, and it is shown here for the sake of completeness;
\item[(B)] The linear sigma model at finite temperature coupled to the Polyakov loop. At this point the effects of the magnetic field were not taken into account yet and we investigated only the effects of the interaction between the two order parameters;
\item[(C)] The linear sigma model in a magnetic background at zero temperature. In this case the Polyakov loop is excluded by construction. The magnetic field manifests itself through vacuum corrections from the quarks;
\item[(D)] The complete scenario: the linear sigma model coupled to the Polyakov loop with thermal corrections in the presence of a magnetic background.
\end{itemize}
In the latter case we also investigated the effects of including vacuum corrections or not. The discussion about the inclusion or not of such terms is still inconclusive, not being clear which treatment is closer to the original theory, QCD. We obtained phase diagrams
in the $(T,B)$ plane
for both cases and verified that the inclusion or not of vacuum corrections is crucial to the structure of the phases, influencing not only the nature of the transitions, as it was first observed in \cite{Boomsma:2009eh}, but also changing the qualitative behavior of the transition lines.

\subsection{$B=0, T\neq 0$, $\varphi_{1,2}=0$}

This case is described by the linear sigma model coupled to quarks. We take into account thermal corrections and disregard coupling to the Polyakov loop~\eq{eq:Phi}. The chiral condensate is the approximate order parameter for the chiral transition, and for low temperatures it is nonzero, indicating symmetry breaking. The value in the vacuum is adjusted to correspond to the pion decay constant, $f_\pi$, as described in section \ref{chiral_lagrangian}. As the temperature is raised the minimum approaches zero, and for high enough $T$ the symmetry is recovered. As the position of the minimum as a function of temperature is continuous, and so is its derivative, the transition is a crossover.

\begin{figure}[!thb]
\vskip 3mm
\begin{center}
\includegraphics[width=85mm,clip=true]{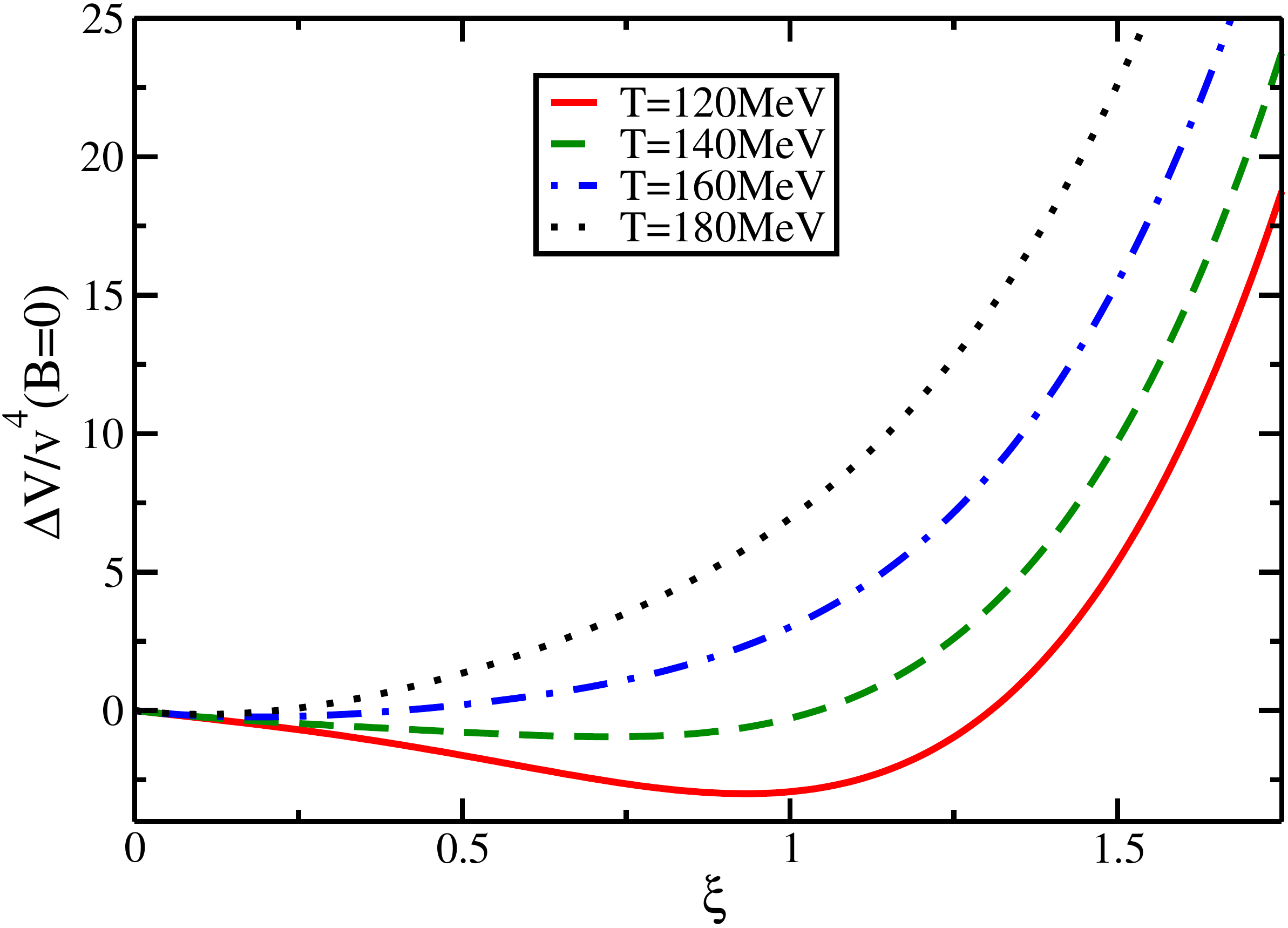}
\end{center}
\vskip -3mm
\caption{Effective potential of the linear sigma model with thermal corrections
for several values of temperature. The transition is a crossover.}
\label{fig:Vcla_thermal_B0}
\end{figure}

\subsection{$B=0$, $T\neq 0$, $\varphi_{1,2}\neq 0$}
\label{sec:B0}

In this subsection we analyze the interaction between the chiral condensate and the Polyakov loop, without the presence of the magnetic field.
The treat the model at finite temperature because of the presence of the Polyakov loop~\eq{eq:Phi} in the formulation of the model.
The expression for the thermal effective potential is similar to Eq.(\ref{eq:V2}):

\beqn
\Omega^{\mathrm{para}}_q = - T
\int \frac{d p^3_z}{(2 \pi)^3} \, W_{B=0}[\omega(p_z,\sigma),\Phi]\,, \quad
\label{eq:V2B0}
\eeqn
where $W_{B=0}$ has an expression similar to Eq.(\ref{eq:W:T}), but the frequency $\omega$ has no dependence on the magnetic field, and is given by $\omega = \sqrt{p^2+m_q^2}$.

\begin{figure}[!thb]
\vskip 3mm
\begin{center}
\includegraphics[width=85mm,clip=true]{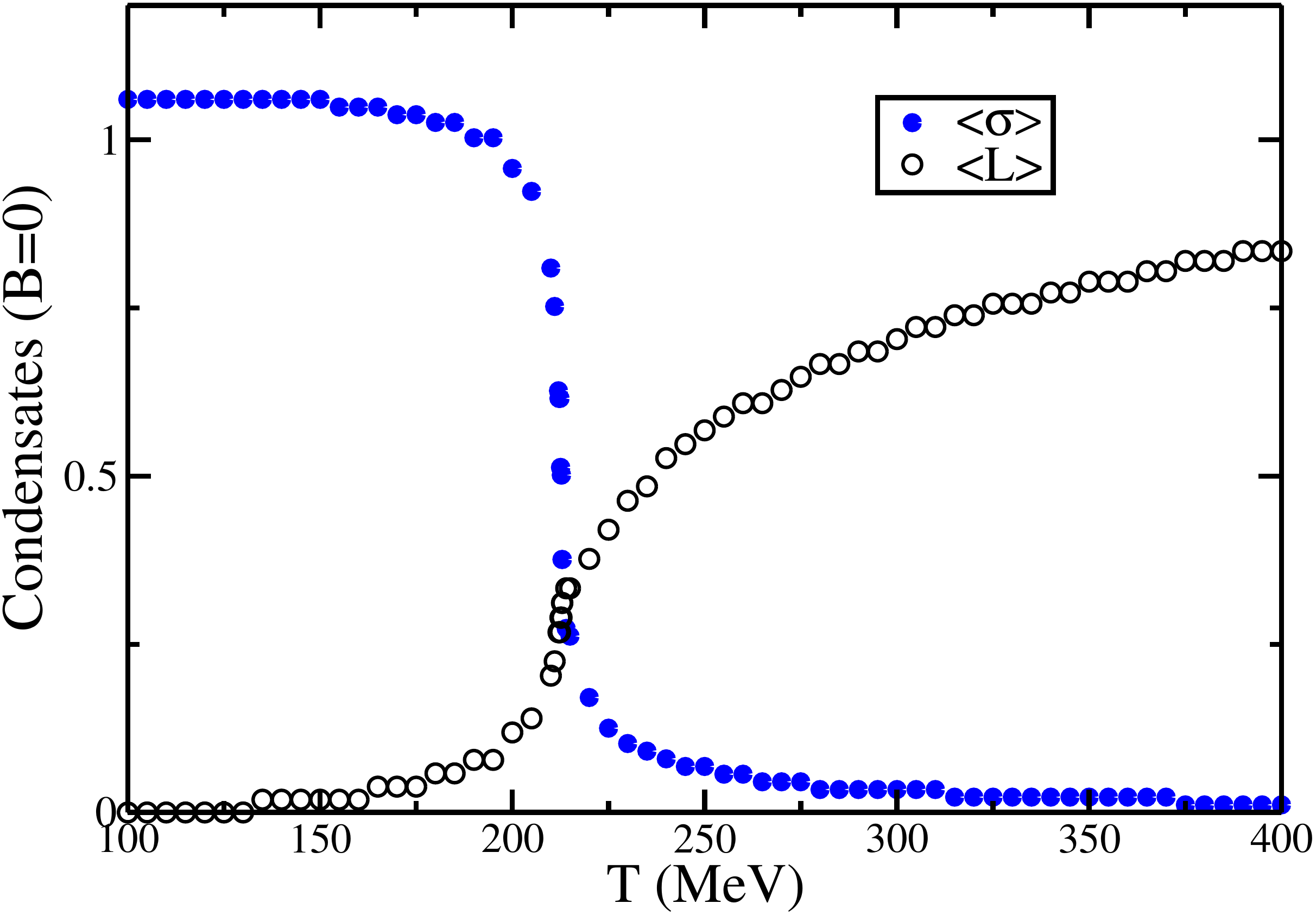}
\end{center}
\vskip -3mm
\caption{
Expectation values
of the order parameters for the chiral and deconfinement transitions as functions of the temperature. The filled circles represent the
$\sigma$--condensate, and the empty circles stand for the expectation value of the Polyakov loop. Both lines are smooth functions of the temperature
indicating the presence of a crossover-type smooth transition.}
\label{fig:B0}
\end{figure}

Due to the interaction, the $\Z_3$ degeneracy is broken and the minimum along the real axis becomes the global one. Both transitions, chiral
and deconfinement, are consistent with a crossover-type transition since the order parameters behave as smooth functions of the temperature,
Fig.~\ref{fig:B0}. The slopes of both order parameters becomes steepest at the same temperature $T_c \simeq 215\,\mbox{MeV}$,
clearly showing the tight relation between the order parameters.

\subsection{$B\neq 0$, $T = 0$}

\begin{figure}[!thb]
\vskip 3mm
\begin{center}
\includegraphics[width=85mm,clip=true]{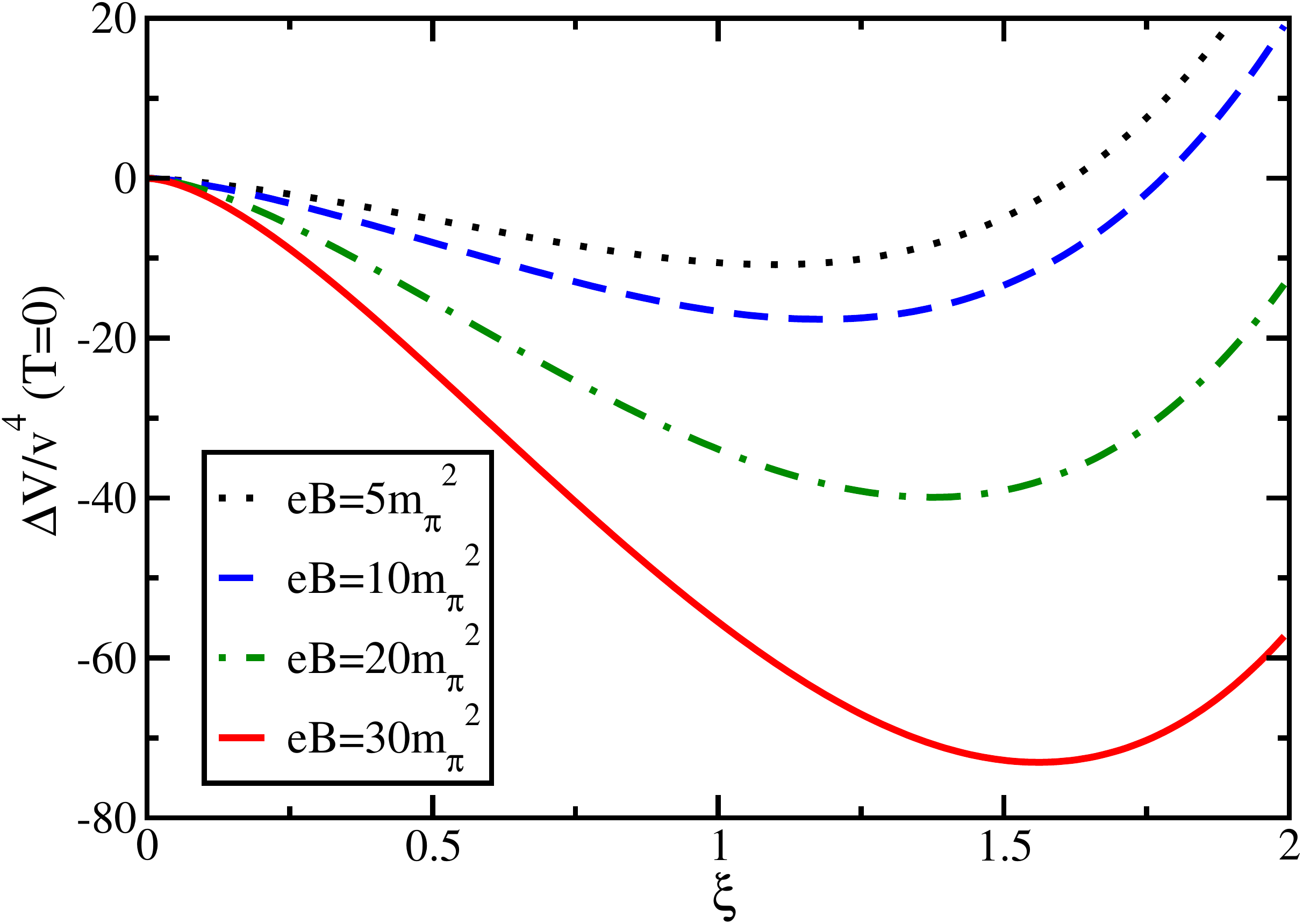}
\end{center}
\vskip -3mm
\caption{
Zero-temperature effective potential~\eq{eq:Veff-dim} with vacuum corrections included is shown
as a function of the chiral variable $\xi=\sigma/v$ for different values of external magnetic field.}
\label{fig:zerotemp_severalB}
\end{figure}

At zero temperature the Polyakov loop is absent by construction. The only pieces of effective potential that contribute are the linear sigma model potential and the vacuum correction from quarks. The magnetic field influencing the system through the latter. We plot in Fig.~\ref{fig:zerotemp_severalB} the effective potential for different values of the external magnetic field. We can see that as the field increases in magnitude the potential becomes deeper and the value of the condensate raises, in accordance with results at zero temperature that indicate an enhancement of chiral symmetry breaking. This effect -- known as magnetic catalysis -- is in accordance with previous result from two of us \cite{Fraga:2008qn}. In Fig.~\ref{fig:zerotemp_severalB} we subtracted the value of the potential with $\xi=0$, in order to have a better comparison for different values of magnetic field, and $\Delta V$ is the resultant potential. This procedure will be applied in all plots of the effective potential along this section.

\subsection{$B\neq 0$, $T\neq 0$, $\phi\neq 0$}

Finally we analyze the full potential. It contains the linear sigma model, the pure gauge potential, thermal corrections including interaction between the two fields, and vacuum corrections from the quarks. The effects of including or not the vacuum terms were also considered and have shown to be crucial for the structure of phases. The relevant potential is given in Eq. (\ref{eq:Veff-dim}). This potential is a function of the parameters $\sigma$,
$\varphi_1$, and $\varphi_2$, where we used the parameterization~\eq{eq:Phi} and set $\varphi_3 = - \varphi_1 - \varphi_2$.
As shown in Fig.~\ref{fig:VL:potential}, for pure gauge the potential -- written in terms of $L$ and $L^*$ -- has three degenerated minima. It is known that in the presence of quarks this degeneracy is broken and the global minimum will be in the direction of $Re[L]$.  Since we are interested only in  the minima of the potential, we can simplify our analysis by setting $Im[L]$ to be zero.

\begin{figure}[!thb]
\vskip 3mm
\begin{center}
\includegraphics[width=85mm,clip=true]{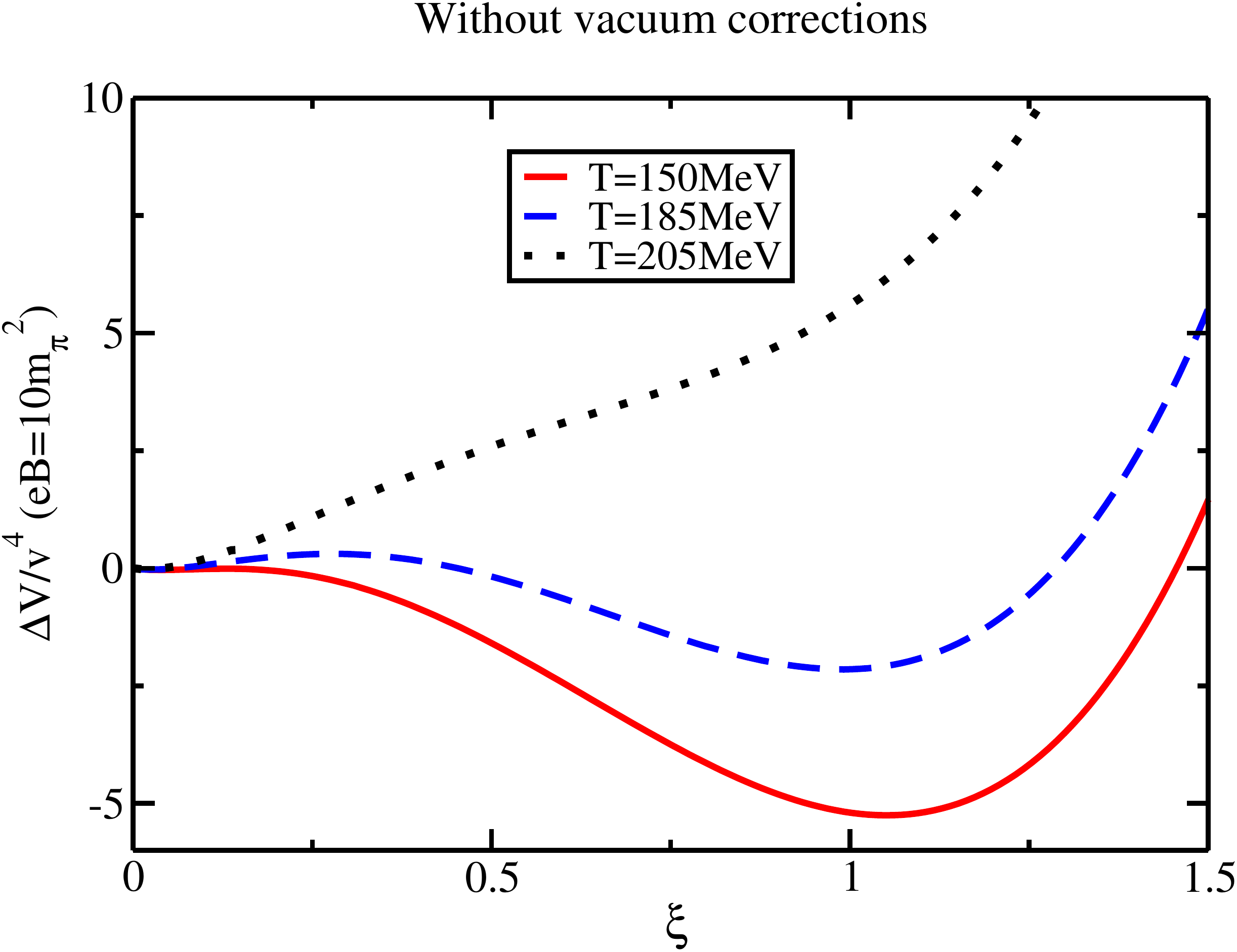}
\end{center}
\vskip -3mm
\caption{Complete effective potential without vacuum corrections from quarks as a function of $\xi=\sigma/v$.
The value of $\varphi$ was chosen to take the minimum for each temperature.
The barrier between the two minima indicates a first-order chiral transition.}
\label{fig:B10pot_nonvacuum}
\end{figure}

\begin{figure}[!thb]
\vskip 3mm
\begin{center}
\includegraphics[width=85mm,clip=true]{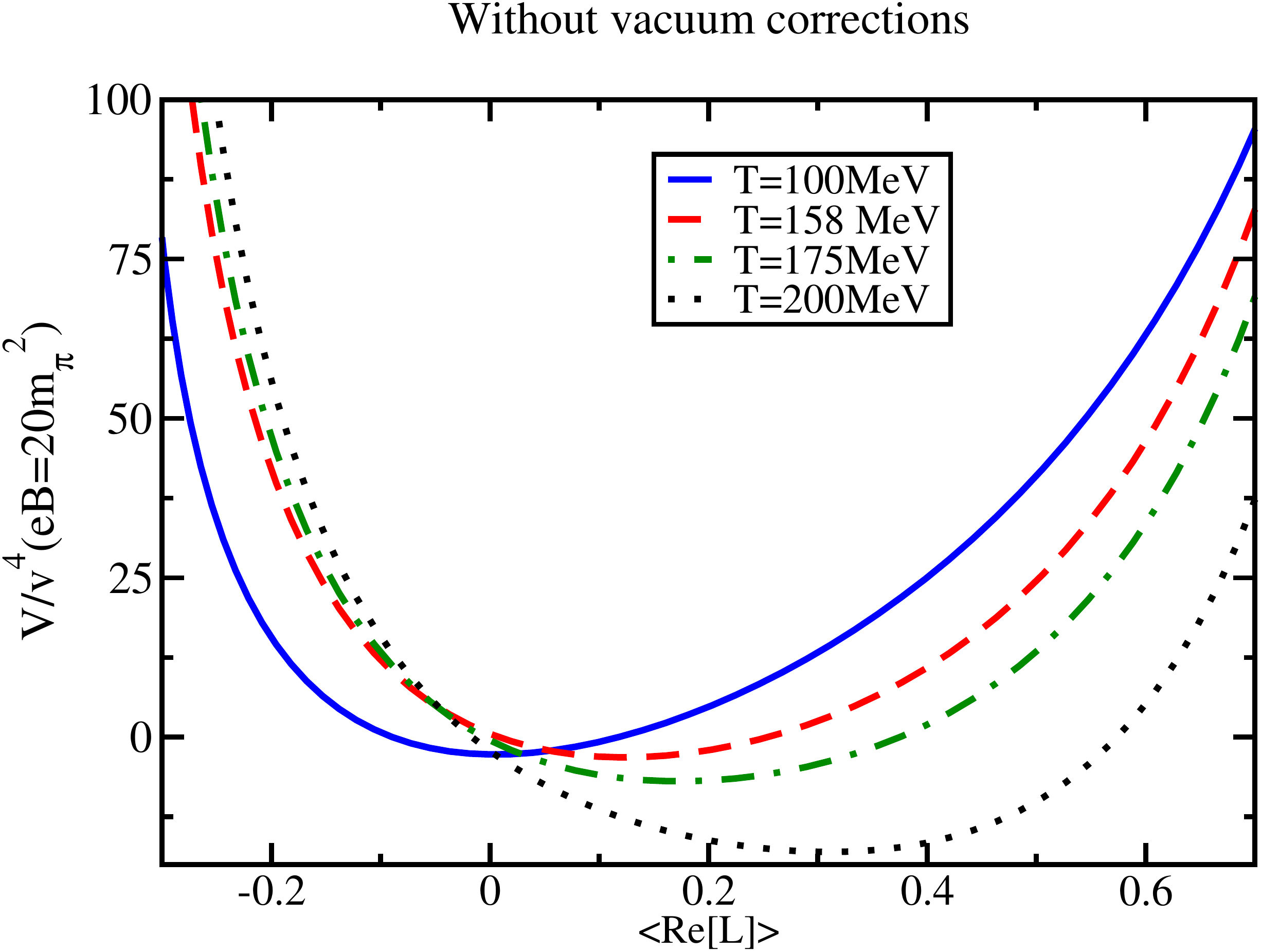}
\end{center}
\vskip -3mm
\caption{Complete effective potential without vacuum corrections from quarks as a function of $Re[L]$. The value of $\sigma$ was chosen to be the minimum for each temperature. The different curves are then slices with different $\sigma$, representing a first order transition,
even if there is no apparent barrier in the plot
(the first order jump in the chiral condensate makes the expectation value of the Polyakov loop discontinuous as well).}
\label{fig:B20L_nonvacuum}
\end{figure}

Let us consider first the potential without vacuum corrections. We evaluate it for different values of magnetic field and temperature. For each set of values of $T$ and $B$ we minimize the potential with respect to $\sigma$ and
$\varphi_1$.
Then we use this value of $\langle \varphi_1\rangle$
to plot the effective potential as a function of $\sigma$ only. Figure~\ref{fig:B10pot_nonvacuum} illustrated the evolution of the chiral transition at the magnetic field $e B = 10m_\pi^2$. Initially the condensate has a non-zero expectation value and this value decreases as the temperature increases. Then, a local minimum appears for $\sigma=0$, and above a certain critical temperature it becomes the global minimum. The barrier between the two minima causes a discontinuity in the expectation value of $\sigma$, pointing out to a first order character of the chiral transition, in accordance with \cite{Fraga:2008qn}. In the same way as in the case $B\neq 0$, $T=0$, the potential becomes deeper for high values of the magnetic field. However, the thermal corrections, responsible for restoring chiral symmetry, also acquire a higher magnitude, bringing the chiral critical temperature down.

In Fig.~\ref{fig:B20L_nonvacuum} we fix $\sigma$ at its minimum for a set of values of $T$ and $B$, and plot the potential as a function of $\varphi_1$. We can then follow the evaluation of the expectation value of the parameter $Re[L]$.
At low temperatures it is zero, indicating confinement, and above a critical temperature it goes asymptotically to
one. At the critical value of the temperature, where the chiral transition occurs, the variable $\langle L \rangle$ also presents a discontinuity,
so that the first-order chiral transition is accompanied by a first-order deconfinement transition for large strengths of the magnetic field.
Thus, if the vacuum corrections are not taken into account, then the magnetic field enhances the order of the phase transition
and the transition changes from the crossover type (realized at $B=0$, Section~\ref{sec:B0}) to the first order type.
\begin{figure}[!thb]
\vskip 3mm
\begin{center}
\includegraphics[width=85mm,clip=true]{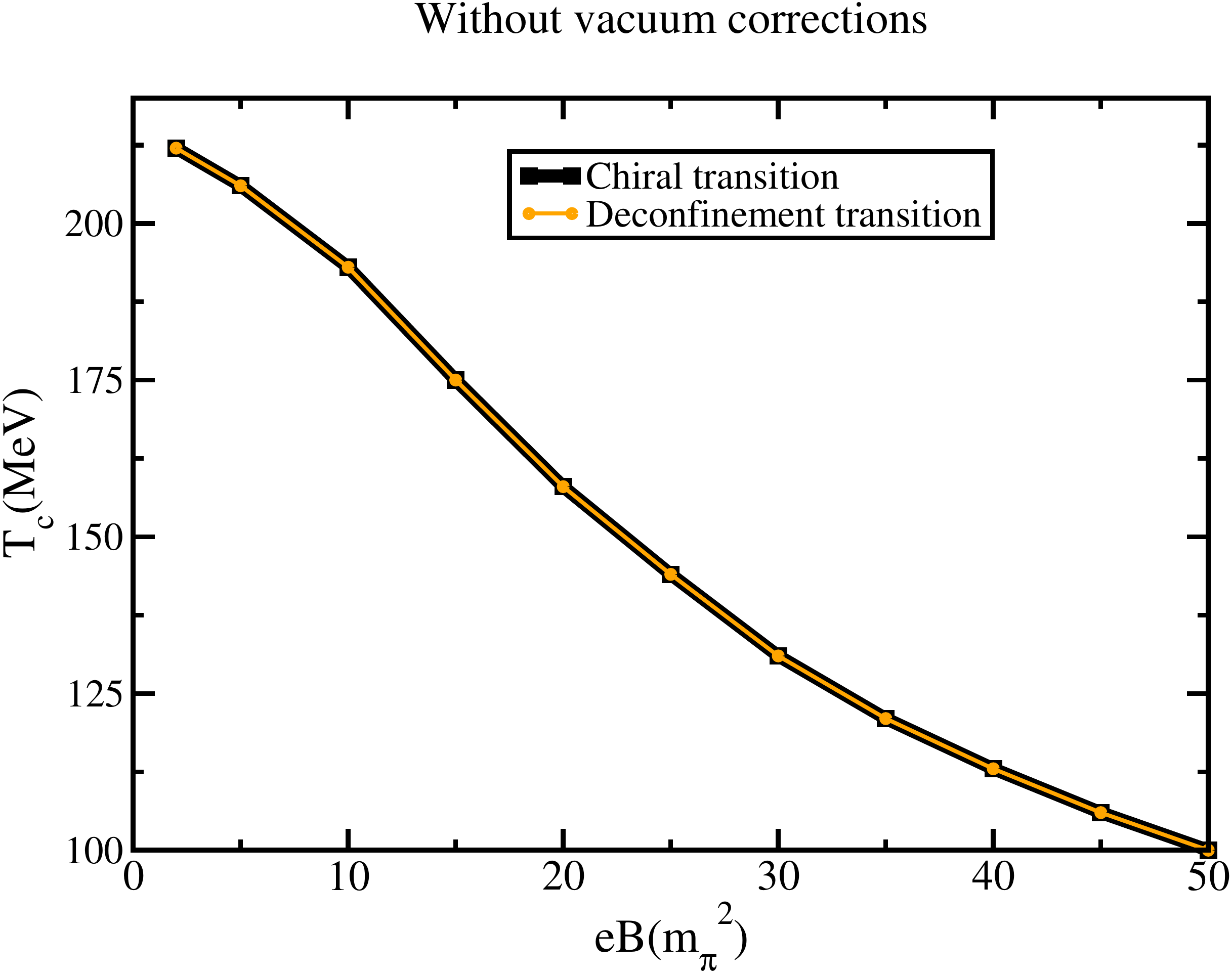}
\end{center}
\vskip -3mm
\caption{
Phase diagram for the complete potential without the vacuum corrections coming from quarks. The chiral (black line) and deconfinement (light red line) transitions happen at the same temperature, leading to a common line. Both critical temperatures decrease with the increase of $B$. The lines mark first-order transitions, except for very small values of $B$, when the transitions become crossovers. The exact value of field for which it happens is not clear from our numerical analysis.}
\label{fig:nonvac_phasedia}
\end{figure}

The phase diagram for the potential described above is shown in Fig.~\ref{fig:nonvac_phasedia} \footnote{
The starting point of the curve -- given by the critical temperature at zero chemical potential -- is somewhat higher
compared to the value obtained in recent numerical calculations of lattice QCD with two flavors of light fermions
(see, for instance, the review~\cite{ref:Karsch}). Generally, effective models like our \M\ or the PNJL model
give approximate answers for particular quantities while being able to predict a general picture at a very good
qualitative level.}. The critical temperature for the chiral and deconfinement transitions are the same for different values of magnetic field, resulting in two phases: a confined phase with broken chiral symmetry, and a deconfined phase with restored chiral symmetry. The common critical
line of these phase transitions is of first order for all values of the magnetic field $B$ except for very small values of $B$. As the magnetic
field decreases, the transition turns back into a crossover reproducing the physical situation with $B=0$ that was discussed around Fig.~\ref{fig:B0}.

\begin{figure}[!thb]
\vskip 3mm
\begin{center}
\includegraphics[width=85mm,clip=true]{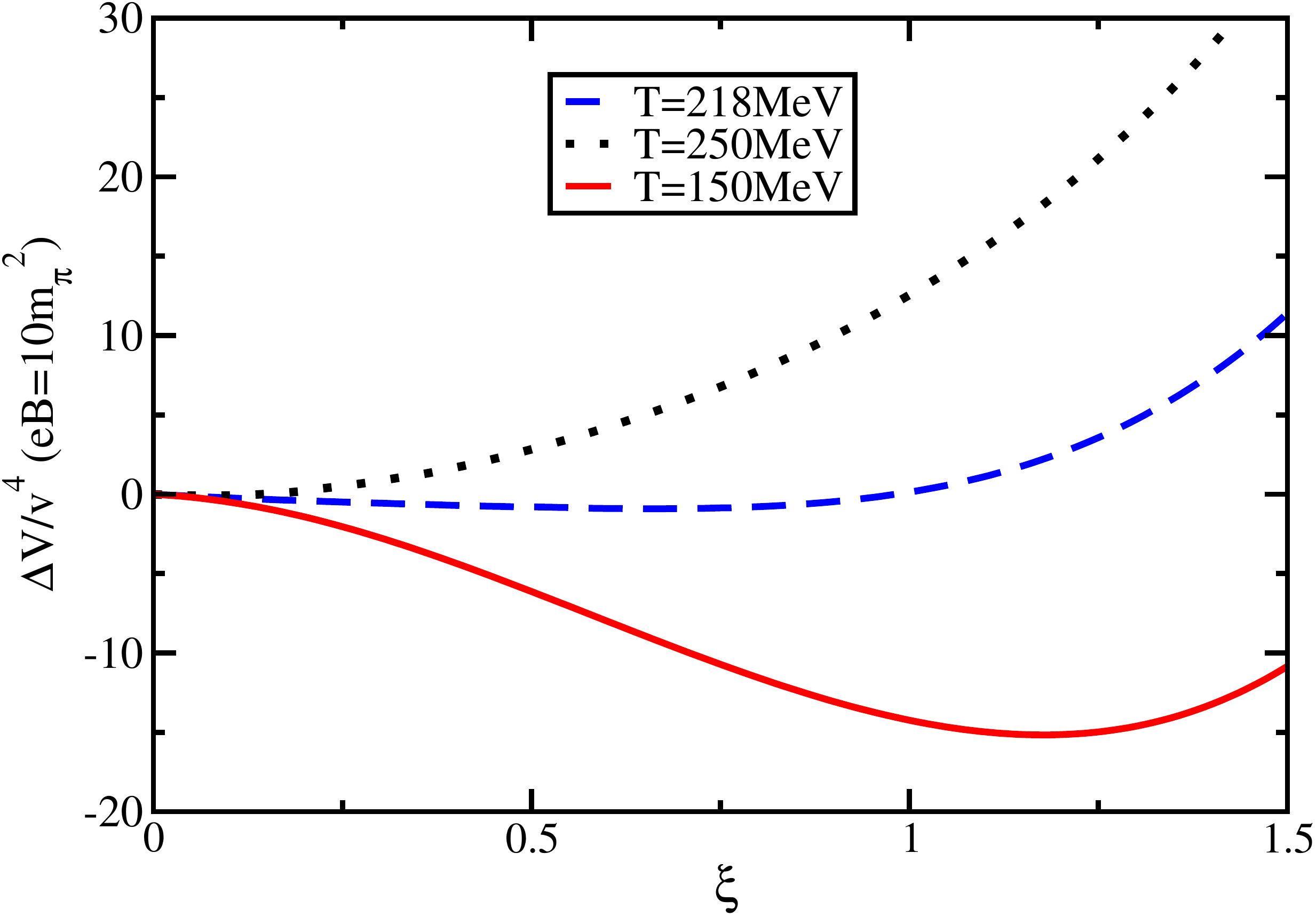}
\end{center}
\vskip -3mm
\caption{
Complete effective potential with vacuum corrections~\eq{eq:Veff-dim} coming from quarks as a function of $\xi=\sigma/v$.
The transition is a crossover.}
\label{fig:B10fullpot}
\end{figure}

\begin{figure}[!thb]
\vskip 3mm
\begin{center}
\includegraphics[width=85mm,clip=true]{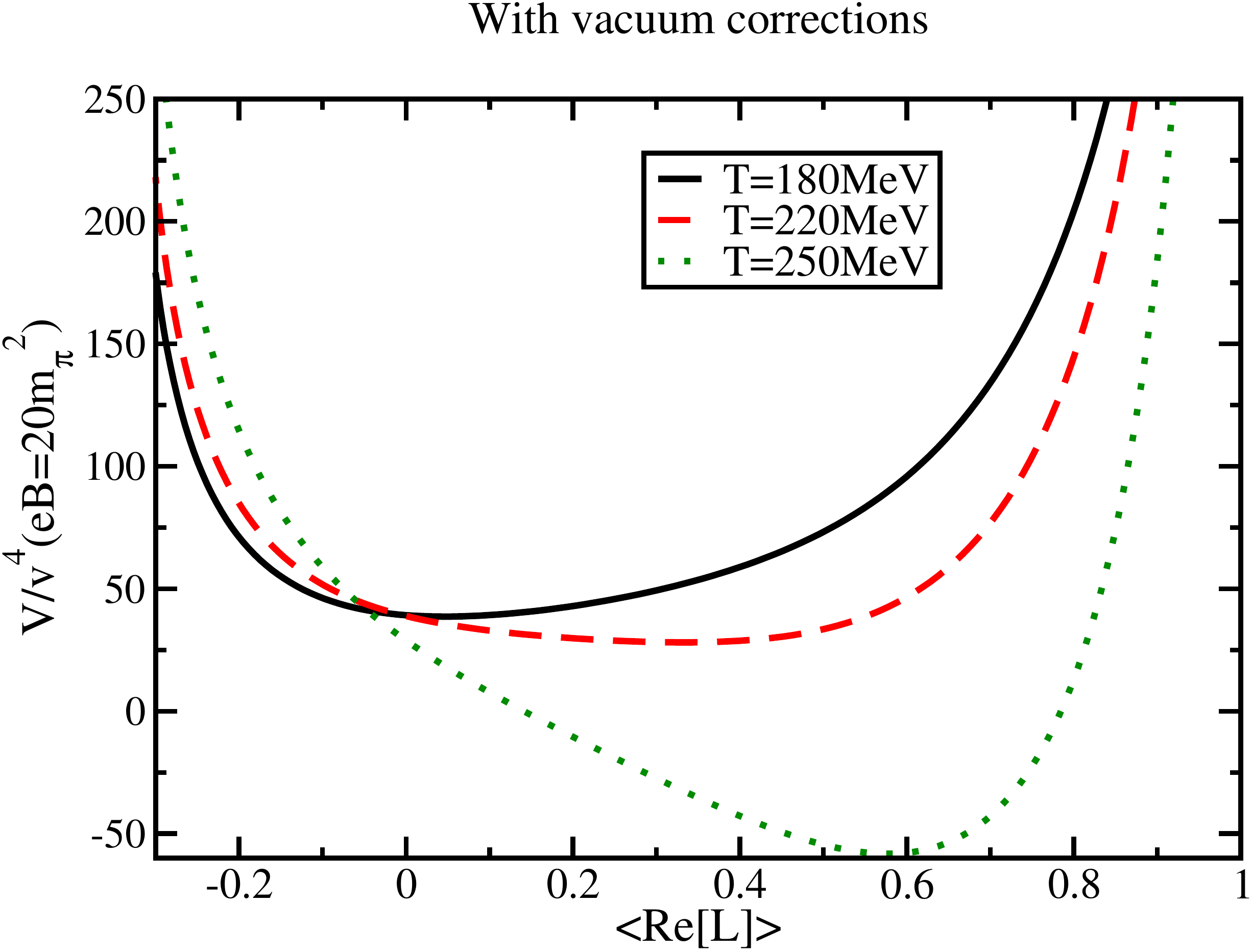}
\end{center}
\vskip -3mm
\caption{Complete effective potential with vacuum corrections from quarks as a function of $Re[L]$ for $B=20m_\pi^2$ and several temperatures.}
\label{fig:B20L_vacuum}
\end{figure}

\begin{figure}[!thb]
\vskip 3mm
\begin{center}
\includegraphics[width=85mm,clip=true]{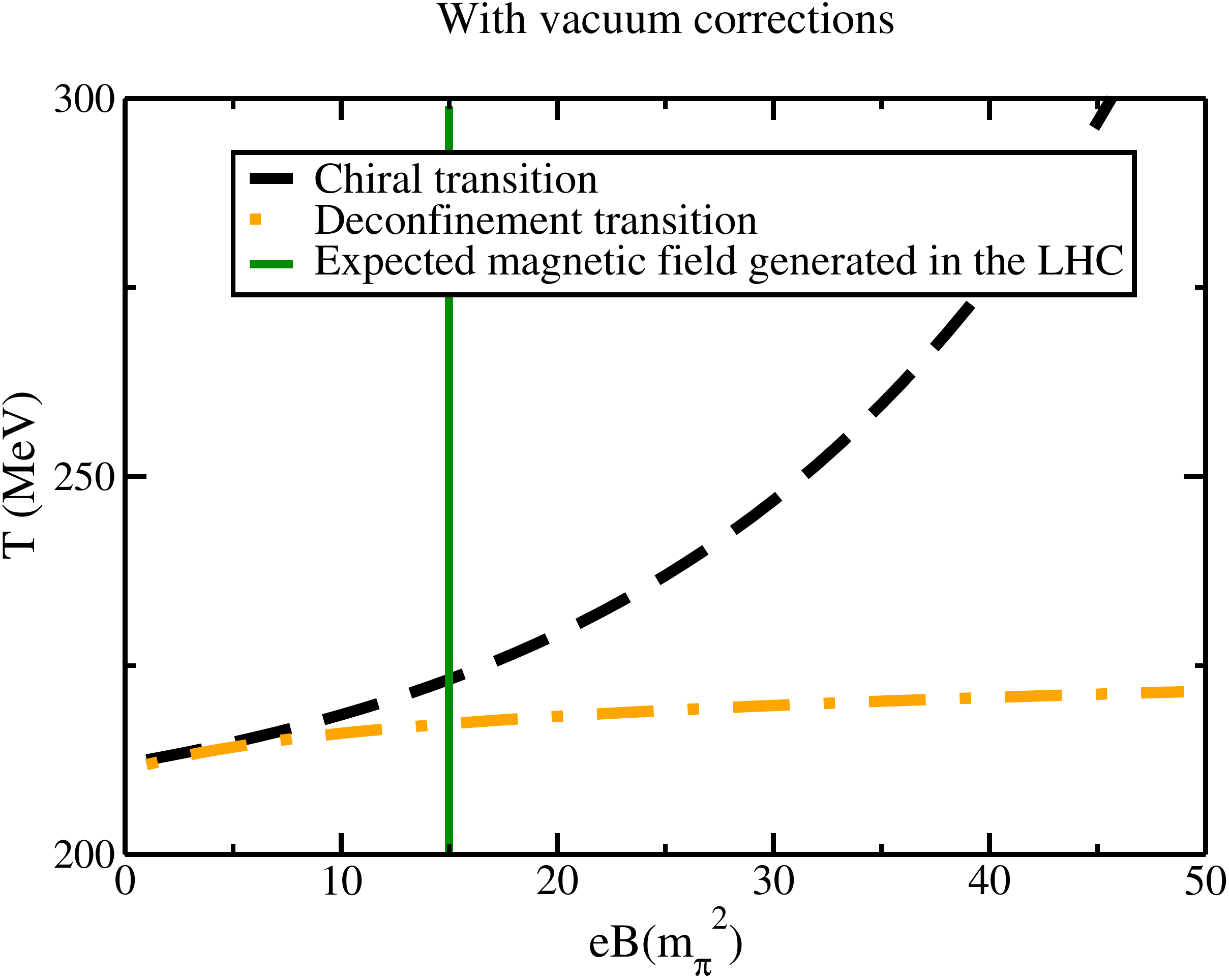}
\end{center}
\vskip -3mm
\caption{Phase diagram for the complete potential with the vacuum corrections from quarks.
The critical temperatures of the deconfinement (the dash-dotted line) and chiral (the dashed line) transition coincide at $B=0$ and split
at higher values of the magnetic field. A deconfined phase with broken chiral symmetry appears.
The green solid vertical line is the
magnitude of the magnetic field that expected to be realized at heavy-ion collisions at the LHC~\cite{Skokov:2009qp}.}
\label{fig:vac_phasedia}
\end{figure}

The inclusion of vacuum corrections from quarks changes the phase diagram dramatically.
Following the same procedure describe above, we plot the potential as a function of $\sigma$ in Fig.~\ref{fig:B10fullpot}
and as a function of $Re[L]$ in Fig.~\ref{fig:B20L_vacuum}.
The corresponding phase diagram is shown in Fig.~\ref{fig:vac_phasedia}. The vertical green line indicates the value of magnetic field
that is expected to be reached in heavy-ion collisions in the ALICE experiment at the LHC~\cite{Skokov:2009qp}.

The presence of the quark vacuum term changes the order of the chiral transition, turning it into a crossover. In the case without the vacuum correction, the transition was of first order due to thermal effects. The thermal contribution generates a minimum at $\xi=0$ being rather sensitive to the value of the magnetic field. In the present case, however, the vacuum term has a magnitude that is larger compared to the thermal piece. Thus, the vacuum contribution dominates the position of the minimum, and it is the vacuum part that is responsible for driving the transition to a crossover. Moreover, the chiral-violating property of the vacuum quark contribution overwhelms the chiral-restoration tendency of the thermal part so that the critical temperature of the chiral transition -- calculated by taking into account both vacuum and thermal effects -- becomes an increasing function of the external magnetic field.

It is interesting to recall that the thermal contribution of quarks tends to destroy the confinement at strong magnetic fields, Fig.~\ref{fig:nonvac_phasedia}, contrary to the confinement-enhancing property of the vacuum contribution.
This effect is due to the fact that the vacuum contribution makes the value of the field $\sigma$ larger (as one can see in Fig.~\ref{fig:zerotemp_severalB}),
suppressing the $\Z_{3}$ violating paramagnetic contribution~\eq{eq:strong:field} at large enough magnetic fields. Thus, large magnetic fields
tend to enhance the confinement property of the vacuum.
As in the case of the chiral transition, the vacuum contribution exceeds its thermal counterpart so that
the deconfinement transition temperature appears to be a weakly increasing function of the magnetic field, Fig.~\ref{fig:vac_phasedia}.
The confinement transition is also a crossover.

The lines of the confinement and chiral transitions coincide for small values of $B$, and go apart as the field increases in magnitude.
Results from lattice QCD indicate that in the absence of the magnetic field the deconfinement transition and chiral symmetry restoration
happen in the same narrow temperature interval~\cite{ref:Karsch:coincide}.
Our calculations indicate that the presence of a strong magnetic field should inevitably split these transitions in this case.

The deconfinement transition occurs at lower temperatures compared to the temperature of chiral restoration. Thus, in a strong magnetic field a deconfined phase with broken chiral symmetry should appear. The splitting between these transitions becomes wider with the increase of the magnetic field $B$. Some other approaches~\cite{Agasian:2008tb} -- that were not taking into account vacuum corrections -- demonstrate that (i) the deconfinement temperature drops down as the magnetic field $B$ increases; and that (ii) at some critical value of the magnetic field, $e B \sim 24 m_\pi^2$, color confinement may disappear as illustrated in Fig.~\ref{fig:expected}. On the contrary, we show that the inclusion of the mentioned vacuum corrections make the critical deconfining temperature to be an increasing function of the magnitude of the magnetic field up to $e B \sim 50 m_\pi^2$. We have not found a signature of a critical magnetic field that would lead to the magnetic-field-induced deconfining transition in either scenario, with and without vacuum corrections.

\section{Conclusion}
\label{sec:conclusion}

In this paper we studied the influence of a strong magnetic field background on confining
and chiral properties of QCD, using the linear sigma model coupled simultaneously
to quarks and to the Polyakov loop (\M). This model associates electromagnetic, chiral and
the confining degrees of freedom in a natural way.

In the confining sector, the strong magnetic field affects the expectation value of the Polyakov loop,
which is an approximate order parameter for the confinement--deconfinement phase transition at finite temperature.
We found that the contribution from the quarks to the Polyakov-loop potential has three features:
\begin{enumerate}
\item the presence of the magnetic field breaks the global $\Z_3$ symmetry and makes the Polyakov loop real-valued
(this effect is seen in the Polyakov loop potential, Fig.~\ref{fig:VL:potential:para});
\item the thermal contribution from quarks tends to destroy the confinement phase by increasing the expectation value of
the Polyakov loop;
\item on the contrary, the vacuum quark contribution tends to restore the confining phase by lowering the expectation value
of the Polyakov loop.
\end{enumerate}

The vacuum correction from quarks has a crucial impact on the phase structure. If one ignores
the vacuum contribution from the quarks, then one finds that the confinement and chiral phase transition lines coincide, Fig.~\ref{fig:nonvac_phasedia}, and in this case the increasing magnetic field lowers the common chiral-confinement
transition temperature. However, if one includes the vacuum contribution, then the confinement and chiral transition lines split, and
both chiral and deconfining critical temperatures become increasing functions of the magnetic field, Fig.~\ref{fig:vac_phasedia}.
The vacuum contribution from the quarks affect drastically the chiral sector as well.

Our calculations also show that the vacuum contribution seems to soften the order of the phase transition: the first-order phase
transition -- which would be realized in the absence of the vacuum contribution -- becomes a smooth crossover
in the system with vacuum quark loops included.

It is important to stress that in a strong magnetic field a deconfined phase with broken chiral symmetry will appear in the scenario with quark vacuum corrections.
The splitting of the confinement and chiral transitions -- as shown in Fig.~\ref{fig:vac_phasedia} -- may be substantial
at a steady magnetic field of the magnitude that is expected to be realized in heavy-ion collisions.
For example, in the ALICE experiment at the LHC facility, the magnetic field may peak around the value
$e B \simeq 15 m_\pi^2$~\cite{Skokov:2009qp}. We find (see Fig.~\ref{fig:vac_phasedia}) that the splitting
between the critical temperatures of the confinement and chiral crossovers in a {\it constant} field of the typical
LHC magnitude may reach the noticeable value
$$(T^{\mathrm{chiral}}_c - T^{\mathrm{deconf}}_c)_{\mathrm{LHC}} \simeq 10\,\mbox{MeV}\,.$$
\vskip 5mm

\vspace*{0.2cm}

Deciding whether quark vacuum contributions should be included or not is, unfortunately, an open issue yet. Lattice simulations of QCD with dynamical fermions in the presence of a magnetic field could certainly shed some light onto this matter. Recent lattice results seem to favor the scenario that takes into account the vacuum 
contributions~\cite{delia}. 

\vspace*{0.2cm}
{\it Note added.} After this paper was submitted to the journal, a new preprint on a similar subject has appeared~\cite{Gatto:2010qs}.
The authors have found a splitting of the deconfinement and chiral transitions in the magnetic field background in the framework of 
a Polyakov-loop extended Nambu-Jona Lasinio model with eight-quark interactions. This independent result is an agreement with 
our conclusion on splitting of these transitions in the linear $\sigma$-model (with the vacuum corrections included).

\acknowledgments
A.J.M. is especially grateful to J. Boomsma for intense and
valuable discussions. The authors are also indebted to
D. Boer, M. D'Elia, K. Fukushima and D. Kharzeev.
A.J.M. and E.S.F. thank the LMPT at Universit\'e de
Tours, and E.S.F is grateful to the members of the theory
group at Vrije Universiteit Amsterdam, where part
of this work has been done, for their kind hospitality.
The work was partially supported by CAPES/COFECUB project number 663/10.
The work of E.S.F. and A.J.M. is partially supported by
CAPES, CNPq, FAPERJ, and FUJB/UFRJ.
The work of M.N.C. is partially supported by the French Agence Nationale 
de la Recherche project ANR-09-JCJC ``HYPERMAG''.

\end{document}